\documentclass[aps,prb,twocolumn,superscriptaddress,longbibliography]{revtex4-1}

\usepackage{graphicx} 
\usepackage{amsmath}
\usepackage{amsthm} %
\usepackage{amssymb}%
\usepackage{amsfonts}
\usepackage{amssymb}
\usepackage{epstopdf}
\usepackage{makeidx}
\usepackage{epsfig}
\usepackage{amsfonts}
\usepackage{bm}
\usepackage{color}
\usepackage{xcolor}
\usepackage{bbold}
\usepackage{afterpage}
\usepackage{empheq}
\usepackage{hyperref}
\usepackage{booktabs}
\usepackage{subfigure}
\usepackage{makecell}
\usepackage{braket}
\usepackage{tikz}
\usepackage{amsmath}
\usetikzlibrary{3d, calc, decorations.markings}
\usetikzlibrary{angles, quotes}
\usepackage{pgfplots}
\usepackage{bm}
\pgfplotsset{compat=1.18}
\newcommand{\be}{\begin{equation}}
\newcommand{\ee}{\end{equation}}
\newcommand{\ba}{\begin{align}}
\newcommand{\ea}{\end{align}}

\begin{document}


\title{Orbital Hall response of spin-1 fermions with a curved central band}

\author{J. H. Warnes}
\affiliation{Departamento Acad\^emico de F\'isica, Universidade Tecnol\'ogica Federal do Paran\'a, 86812-460, Apucarana, Paran\'a, Brazil}
\author{Dayana Joy}
\affiliation{Department of Physics, School of Engineering and Sciences, SRM University AP, Amaravati, 522240, India}
\author{Pankaj Bhalla}
\affiliation{Department of Physics, School of Engineering and Sciences, SRM University AP, Amaravati, 522240, India}
\author{Rhonald Burgos Atencia}
\affiliation{Direcci\'on de Ciencias B\'asicas, Universidad Tecnol\'ogica de Bol\'ivar, Cartagena, Colombia}

\begin{abstract}
We investigate the orbital Hall effect in a three-band system described by a low-energy Hamiltonian containing both linear and quadratic momentum-dependent contributions. The linear term lifts the degeneracy between the dispersive bands, providing a minimal setting to study orbital transport in a multiband system with a curved central band. Within a quantum kinetic approach, we derive the intrinsic orbital Hall response and analyze its microscopic origin in terms of interband coherence induced by an external electric field. We show that the orbital Hall conductivity arises from the interplay between the momentum-dependent orbital character of the eigenstates and transitions between different bands. The spin-1 structure provides a particularly transparent platform in which the contributions associated with different interband channels can be identified separately.
We further show that the differential orbital Hall response provides a direct probe of the low-energy structure, exhibiting distinct scaling singularities associated with the linear and quadratic bands, respectively, while remaining independent of the ultraviolet cutoff. Our results establish a connection between the multiband quantum geometry of spin-1 fermions and their transverse orbital response, and provide a simple framework for investigating orbital transport beyond two-band models.
\end{abstract}

\maketitle

\section{Introduction}

The study of orbital angular momentum transport in condensed matter systems has attracted considerable attention in recent years, driven both by its fundamental interest and by its potential applications in
information processing and spin-orbitronic devices \cite{GoDongwook2017,GoDongwook2021,Kim2022,Ando2025,WangPing2025,Yesudas2026}. 
Traditionally, spintronics has focused primarily on the generation, manipulation, and detection of spin currents, with the electron spin regarded as the main carrier of angular
momentum in solids \cite{Murakami2003,DasSarma2004,Murakami2004,Kato2004}. 
However, electrons also possess an orbital degree of freedom that can participate actively in transport phenomena \cite{Sundaram1999,Haldane2004,Chang2008,DiXiao2010,ZhongShudan2016,Pozo2023}. 
Recent theoretical and experimental developments have revealed that orbital angular 
momentum can generate sizable transverse currents and can play an important role in the transfer of angular momentum in a broad variety of materials \cite{GoDongwook2020,GiacomoSala2022,Cysne2025}. 
These advances have established orbital transport as a complementary
route to conventional spin-based mechanisms and have motivated an increasing effort to understand the microscopic and geometrical origins of orbital currents in multiband electronic systems.

Among the phenomena associated with orbital angular momentum transport, the orbital Hall effect (OHE) has emerged as one of the most prominent examples \cite{BurgosAtencia2024,Cullen2025,Cullen2026}. 
In the OHE, an applied electric field or, equivalently, a longitudinal charge current induces a transverse flow associated with the orbital magnetic 
moment (OMM) of the electronic states \cite{PezoArmando2023II,Jo2024}. In close analogy with the spin Hall effect, this transverse response can arise even in the absence of a net
equilibrium orbital polarization and provides a mechanism for generating and transporting angular momentum through a solid. 

Microscopically, the OHE originates from the interplay between the orbital 
character of the electronic states and their momentum-dependent evolution throughout the band structure \cite{GoDongwook2018,PezoArmando2022,CysneTarik2022}. 
As a consequence, interband coherence and the geometric properties of Bloch states can play a central role
in determining the magnitude and structure of the orbital Hall response \cite{LiuHong2024,HongLiu2025,Cysne2026}. 
The OHE is therefore intrinsically connected to the multiband nature of electronic systems and provides a particularly useful setting for exploring
how orbital degrees of freedom, band structure and quantum geometry \cite{Provost1980,Jiang2025,YuJiabin2025,Verma2026}  combine to produce transverse transport phenomena. 

Theoretical and experimental studies have shown that the OHE can arise even in materials with 
weak spin-orbit coupling, indicating that strong relativistic interactions are not
a necessary condition for the generation of transverse orbital currents \cite{JoDaegeun2018,Choi2023,Busch2023,Marui2023,Das2025,Rang2025}.  
Recent experiments have further provided evidence for long-range transport of orbital angular momentum in metallic heterostructures~\cite{Seifert2023}.
This feature distinguishes orbital transport from its spin counterpart and considerably broadens the range of material
in which Hall-like angular momentum transport can occur. Moreover, in systems with finite spin-orbit coupling, orbital currents can be converted into spin currents \cite{Lee2021,Zheng2024,HuChen2025,Ding2025,Chen2025,Chen2026,LiuPengfei2026,Ning2026}, providing a possible microscopic 
mechanism contributing to the spin Hall effect (SHE).

From this perspective, the OHE can represent a primary response of the electronic orbital degree of freedom, while spin-orbit coupling provides
a channel through which orbital angular momentum is transferred to the spin sector. Understanding the OHE is therefore relevant not only for orbital transport itself, but also for clarifying the microscopic mechanism underlying spin transport in solids. 

Pioneering studies of the OHE were developed in spin-$1$ systems with band degeneracies, where the orbital dynamics arises from the coherent mixing between different orbital states. 
In particular, Bernevig \textit{et al.} \cite{Bernevig2005} demonstrated that an applied electric field can generate a transverse orbital current in a spin-$1$ model describing the $p$-bands of the hole-doped silicon. In this model, the three $p$-orbital degrees of freedom transform as a spin-$1$ representation and the orbital Hall response emerges in the presence of degenerate bands. However, the consequences of this mechanism in a spin-1 system with a curved central band remain much less explored. In this regard, it is particularly interesting to investigate how central-band curvature modifies the microscopic origin of the transverse orbital response.

The absence of degeneracy changes the structure of the interband process
responsible for orbital transport and raises the question of whether the OHE persists and which geometrical properties of the electronic states govern it. In this work, we address these questions by studying the OHE in 
a spin-1 system with a curved central band, where a momentum-linear
coupling produces two linearly dispersing bands.

\section{Theoretical model and quantum kinetic equation formalism}

\subsection{Model Hamiltonian and band structure}

The system considered here is described by a Hamiltonian of the form
$H=H_0+H_E$, where $H_0$ is the band Hamiltonian and
$H_E=e\bm E\cdot\bm r$ is the driving term. We consider a uniform and
time-independent electric field,
$\bm E(\bm r,t)=\bm E_0$, where $\bm r$ is the single-particle position
operator. For the band Hamiltonian, we consider a spin-$1$ system
described by~\cite{Kikuchi2025,Kikuchi2026}

\begin{equation}
H_0(\bm k)
=
-ck^2 I_0
+
\hbar v_F\bm S\cdot\bm k
+
c(\bm S\cdot\bm k)^2,
\label{eq:H0}
\end{equation}
where $I_0$ is the $3\times 3$ identity matrix,
$\bm S=(S_x,S_y,S_z)$ denotes the spin-$1$ matrices,
$\bm k$ is the crystal wave vector with $k=|\bm k|$,
$v_F$ is the Fermi velocity, and $c$ controls the quadratic correction
to the band structure. The first term, $-ck^2I_0$, provides a
quadratic energy shift common to all three bands, while the last term,
$c(\bm S\cdot\bm k)^2$, introduces a quadratic correction associated
with the spin-$1$ structure. Their combined effect is to endow the otherwise
flat central band of the linear spin-$1$ model with a finite
curvature~\cite{Mizoguchi2021}. 
Spin-$1$ systems have previously been considered in the context of charge transport \cite{Wang2017,Burgos2022-I,Kikuchi2022},  but its orbital degree of freedom has been less explored.
Spin-$1$ Hamiltonians have also acquired direct experimental relevance
through the discovery of multifold fermions ~\cite{BarryBradlyn2016,NiZhouliang2020,Ni2020,KaushikSahal2021,QunYang2023,Robredo2024,Balduini2024} in chiral semimetals \cite{Chang2017,Tang2017,Sanchez2019,Bing2020,Ni2021}, where threefold fermions and their characteristic interband
transitions have been observed experimentally~\cite{Rao2019,Xu2020}.

The matrices $S_i$ form the spin-$1$ representation of the SU(2) algebra and obey the commutation relations
$[S_i,S_j]=i\varepsilon_{ijl}S_l$. Defining the helicity operator as
$\hat{h}_0=\hat{\bm k}\cdot\bm S$, one finds that it commutes with the
Hamiltonian, $[\hat{h}_0,H_0]=0$. Therefore, helicity is a conserved quantity
and the eigenstates of $H_0$ can be simultaneously chosen as eigenstates
of $\hat{h}_0$. The energy eigenstates are thus labeled by the helicity quantum
number $m=0,\pm1$, satisfying
\begin{equation}
H_0(\bm k)
|u_{m\bm k}\rangle
=
\varepsilon_{m\bm k}
|u_{m\bm k}\rangle,
\end{equation}
where $|u_{m\bm k}\rangle$ denotes the periodic part of the Bloch state.
The corresponding energy spectrum is
\begin{equation}
\varepsilon_{m\bm k}
=
m\hbar v_F k
+
c(m^2-1)k^2.
\label{eq:spectrum}
\end{equation}
The spectrum therefore consists of two linearly dispersing helicity
bands, $\varepsilon_{\pm}=\pm\hbar v_F k$, and a central band,
$\varepsilon_0=-ck^2$. A remarkable feature of
Eq.~(\ref{eq:spectrum}) is that the two helicity bands remain completely
unaffected by the quadratic correction, whereas the curvature of the
central band is entirely controlled by $c$. Consequently, the threefold
band crossing at $k=0$ is preserved for any value of $c$.

\begin{figure}[t]
\centering
\includegraphics[width=\linewidth]{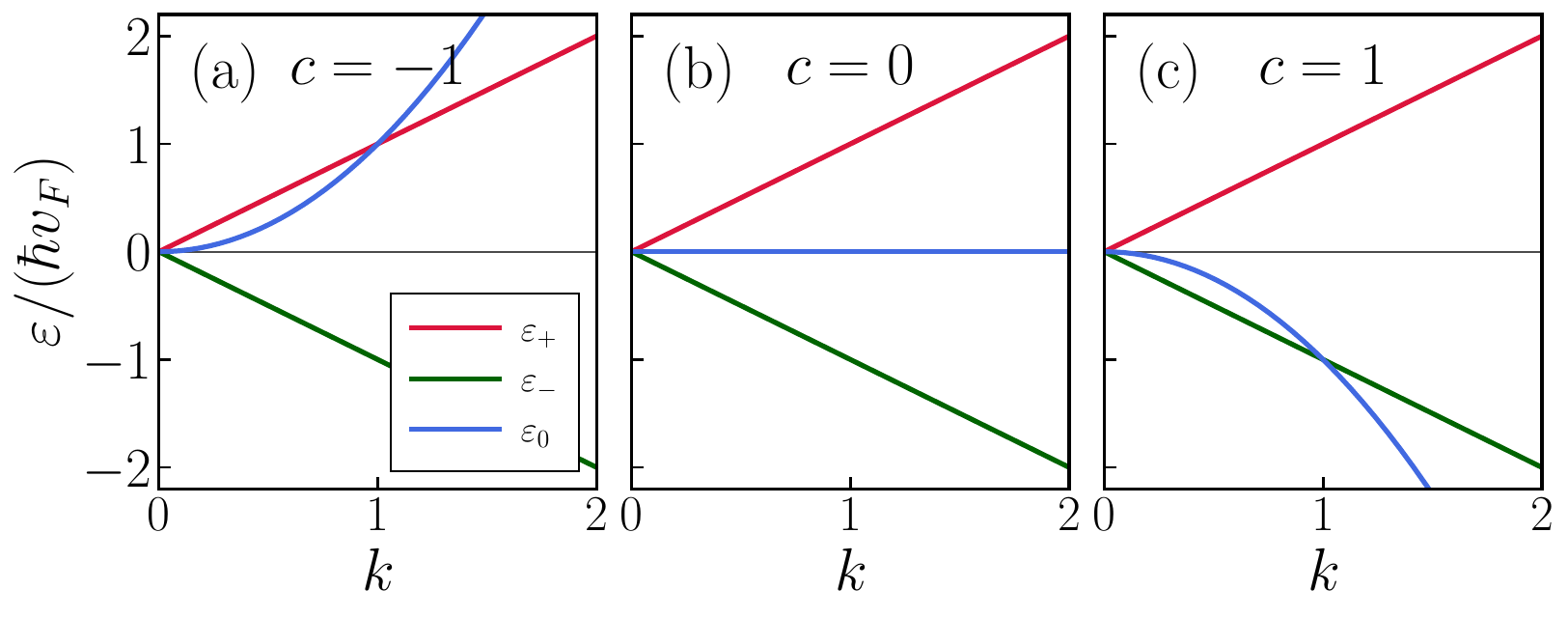}
\caption{ Electronic band structure for (a) $c<0$, (b) $c=0$, and
(c) $c>0$. The linear bands $\varepsilon_{\pm}=\pm\hbar v_F k$ are
independent of $c$, while the central band
$\varepsilon_0=-ck^2$ is upward-opening, flat, or downward-opening,
respectively. The panels correspond to $c=-1$, $0$, and $1$, with
$\hbar=v_F=1$.}
\label{fig:band_structure}
\end{figure}

Figure~\ref{fig:band_structure} illustrates the evolution of the band
structure for three representative values of the quadratic parameter $c$.
%
%
For $c=0$, the central band is completely flat and intersects the two
helicity bands only at the threefold node at $k=0$. For $c<0$, the
central band bends upward and crosses the upper helicity band, whereas
for $c>0$ it bends downward and crosses the lower helicity band. In both
cases, an additional finite-momentum crossing occurs at $k_c=\hbar v_F/|c|$.
%
At $k=k_c$, the central band becomes degenerate with one of the helicity bands. As discussed below, the contribution of the corresponding interband channel to the orbital Hall response vanishes exactly at the crossing because the two degenerate states have identical occupation factors. Nevertheless, the central band remains coupled through interband processes to the opposite-helicity band, whose excitation energy remains finite. This distinction between the two interband channels will play an important
role in determining the orbital Hall response.

The eigenstates of Eq.~(\ref{eq:H0}) are the helicity eigenstates and can
be constructed by performing a rotation of the conventional $S_z$
eigenbasis. Starting from the states $|m\rangle_z$, defined through
$S_z|m\rangle_z=m|m\rangle_z$ ($m=\pm1,0$), the corresponding helicity
eigenstates are obtained as \cite{BurgosAtencia2026}
\begin{equation}
|u_{m\bm k}\rangle
=
e^{-i\phi S_z}
e^{-i\theta S_y}
|m\rangle_z,
\label{eq:helicitystates}
\end{equation}
where $\theta$ and $\phi$ are the polar and azimuthal angles defining the
direction of the momentum,
\begin{equation}
\hat{\bm k}
=
(\sin\theta\cos\phi,\sin\theta\sin\phi,\cos\theta).
\end{equation}
The transformation in Eq.~(\ref{eq:helicitystates}) corresponds to a
rotation of the quantization axis from the $z$ direction to the direction
of $\bm k$. In terms of Euler rotations, the operator
$e^{-i\phi S_z}e^{-i\theta S_y}$ first performs a rotation by $\theta$
about the $y$ axis and subsequently a rotation by $\phi$ about the $z$
axis. Consequently, the resulting states satisfy
\begin{equation}
\hat{h}_0^p|u_{m\bm k}\rangle = m^p|u_{m \bm k}\rangle,
\end{equation}
where \(p\) denotes an integer power of the helicity operator, making explicit that \(|u_{m\bm k}\rangle\) is an eigenstate of \(\hat h_0\) with helicity \(m\). 
An important property of these states is that their momentum dependence
enters exclusively through the direction $\hat{\bm k}$ and is independent
of the quadratic parameter $c$. Therefore, varying $c$ modifies the
energy spectrum while leaving the helicity eigenstates unchanged. This
separation between the energy dispersion and the momentum-space structure
of the eigenstates will be particularly useful below, since the interband
matrix elements entering the orbital magnetic moment and the orbital Hall
response are determined by the momentum dependence of the helicity states.


\subsection{Quantum kinetic equation}

The starting point for the quantum kinetic description of the transport
response is the Liouville equation for the single-particle density
matrix $\rho$. It reads \cite{Culcer2017,Burgos2022},
\begin{equation}
\frac{\partial \rho}{\partial t}
+
\frac{i}{\hbar}[H,\rho]
=
0.
\label{eq:Liouville}
\end{equation}
In the presence of an external electric field, the density matrix can be
decomposed into its equilibrium and nonequilibrium contributions as
\begin{equation}
\rho
=
\rho_0+\rho_E,
\label{eq:density_decomposition}
\end{equation}
where $\rho_0$ is the equilibrium density matrix and $\rho_E$ denotes
the correction induced by the electric field. Since we are interested
in the linear-response regime, throughout this work $\rho_E$ is retained
only to first order in $\bm E$. Substituting Eq.~(\ref{eq:density_decomposition}) into
Eq.~(\ref{eq:Liouville}) and keeping terms up to linear order in the
electric field gives
\begin{equation}
i\hbar
\frac{\partial \rho_E}{\partial t}
=
[H_0,\rho_E]
+
[H_E,\rho_0].
\label{eq:kinetic_general}
\end{equation}
Here we have used the equilibrium condition
$[H_0,\rho_0]=0$ and neglected the term $[H_E,\rho_E]$, which is
second order in the electric field. In this work we consider a uniform and time-independent electric field.
In the stationary regime, $\partial_t\rho_E=0$, and the quantum kinetic
equation reduces to
\begin{equation}
[H_0,\rho_E]
=
-eE_\beta[r_\beta,\rho_0],
\label{eq:kinetic_stationary}
\end{equation}
where $\beta$ denotes a Cartesian index  and summation over repeated Cartesian indices is implied. Projecting Eq.~(\ref{eq:kinetic_stationary}) onto the eigenstates of
$H_0$, one obtains
\begin{equation}
(\varepsilon_{m'\bm k}-\varepsilon_{m\bm k})
\rho_{\bm k,E}^{m'm}
=
-eE_\beta
\langle u_{m'\bm k}|
[r_\beta,\rho_0]
|u_{m\bm k}\rangle,
\label{eq:kinetic_bandbasis}
\end{equation}
where $\rho_{\bm k,E}^{m'm}=\langle u_{m'\bm k}|
\rho_E |u_{m\bm k}\rangle$.
For $m'\neq m$, this equation determines the band off-diagonal part of
the density matrix,
\begin{equation}
\rho_{\bm k, E}^{m'm}
=
-eE_\beta
\frac{
\langle u_{m'\bm k}|
[r_\beta,\rho_0]
|u_{m\bm k}\rangle
}{
\varepsilon_{m'\bm k}-\varepsilon_{m\bm k}
}, \quad m\neq m'.
\label{eq:rhoE_intermediate}
\end{equation}
The off-diagonal components $\rho_{\bm k,E}^{m'm}$ describe the interband
coherence generated by the external electric field and constitute the
intrinsic contribution relevant to the Hall response considered below.

The next step is to evaluate the commutator involving the position
operator. In the Bloch representation, the matrix elements of the
position operator can be written as
\begin{equation}
(r_\beta)^{m'm}
=
i\delta_{m'm}\partial_{k_\beta}
+
\mathcal A_\beta^{m'm}(\bm k),
\label{eq:position_operator}
\end{equation}
where
\begin{equation}
\mathcal A_\beta^{m'm}(\bm k)
=
i
\langle u_{m'\bm k}|
\partial_{k_\beta}
u_{m\bm k}\rangle
\label{eq:Berry_connection}
\end{equation}
is the Berry-connection matrix. In particular, for $m'\neq m$ we find $(r_\beta)^{m'm}=\mathcal A_\beta^{m'm}$.
The off-diagonal Berry connection characterizes the variation of the
Bloch eigenstates in momentum space and determines the geometrical
coupling between different bands. 

The equilibrium density matrix is diagonal in the band basis and can be written as
\begin{equation}
\rho_0
=
\sum_n
n_{F}(\varepsilon_{n\bm k})
|u_{n\bm k}\rangle
\langle u_{n\bm k}|,
\label{eq:rho_equilibrium}
\end{equation}
where $n_{F}(\varepsilon)$ denotes the Fermi--Dirac distribution. Using this spectral decomposition, the off-diagonal matrix elements of the commutator, for $m'\neq m$, become
\begin{align}
\label{eq:commutator_position}
\langle u_{m'\bm k}|
[r_\beta,\rho_0]
|u_{m\bm k}\rangle
&=
\mathcal A_\beta^{m'm}(\bm k)
\left[n_{F}(\varepsilon_{m\bm k})-n_{F}(\varepsilon_{m'\bm k})\right].
\end{align}
Substituting Eq.~(\ref{eq:commutator_position}) into
Eq.~(\ref{eq:rhoE_intermediate}), we finally obtain the off-diagonal density-matrix elements
\begin{align}
\label{Eq:offdiagonal}
\rho_E^{m'm}(\bm k)&=
\frac{
eE_\beta\,
\mathcal A_\beta^{m'm}(\bm k)
}{
\varepsilon_{m'\bm k}-\varepsilon_{m\bm k}
}
\left[
n_{F}(\varepsilon_{m'\bm k})
-
n_{F}(\varepsilon_{m\bm k})
\right]. 
\end{align}
Equation~\eqref{Eq:offdiagonal} describes the intrinsic interband
coherence generated by the electric field. Its structure makes explicit
the two ingredients controlling the response: the geometrical coupling
between different bands, encoded in the off-diagonal Berry connection
$\mathcal A_\beta^{m'm}$, and the corresponding interband energy
separation $\varepsilon_{m'\bm k}-\varepsilon_{m\bm k}$. As shown below,
this electric-field-induced coherence, combined with the interband
structure of the orbital magnetic moment current, gives rise to the intrinsic
orbital Hall response.

\subsection{Intrinsic orbital Hall conductivity}

We now derive the intrinsic orbital Hall conductivity starting from the
interband density-matrix response obtained in Eq.~(\ref{Eq:offdiagonal}).
The central idea is that an external electric field generates interband
quantum coherence, which, when combined with the orbital magnetic moment current operator
carried by the Bloch states, gives rise to a transverse orbital current.
As shown below, the resulting response can be expressed in terms of
intrinsic geometrical quantities associated with the Bloch eigenstates.

The orbital current operator is defined as the symmetrized product of
the orbital angular momentum and the velocity operator,
\begin{equation}
\mathcal J_\alpha^\gamma
=
\frac{1}{2}
\left\{
L_\gamma,v_\alpha
\right\},
\label{eq:orbital_current_definition}
\end{equation}
where $\alpha$ denotes the direction of the current flow and $\gamma$
labels the component of orbital angular momentum being transported.
The orbital angular momentum is related to the orbital magnetic moment $\mathcal M$
through
\begin{equation}
L_\gamma
=
-\frac{\hbar}{g_L\mu_B}
\mathcal M_\gamma,
\label{eq:L_M_relation}
\end{equation}
where $g_L$ is the orbital Land\'e factor and $\mu_B$ is the Bohr
magneton. The orbital current operator can therefore be written as \cite{Bhowal2021}
\begin{equation}
\mathcal J_\alpha^\gamma
=
-\frac{1}{2}
\frac{\hbar}{g_L\mu_B}
\left\{
\mathcal M_\gamma,v_\alpha
\right\}.
\label{eq:orbital_current_M}
\end{equation}
Its matrix elements in the Bloch-state basis are consequently
\begin{equation}
(\mathcal J_\alpha^\gamma)^{mm'}
=
-\frac{1}{2}
\frac{\hbar}{g_L\mu_B}
\sum_n
\left(
\mathcal M_\gamma^{mn}v_\alpha^{nm'}
+
v_\alpha^{mn}\mathcal M_\gamma^{nm'}
\right).
\label{Eq:OCgeneral}
\end{equation}

In the present work, we focus on the semiclassical orbital magnetic
moment associated with an individual nondegenerate band \cite{Sundaram1999}. Within this
description, the OMM is band diagonal,
\begin{equation}
\mathcal M_\gamma^{mm'}
=
\delta_{mm'}\mathcal M_\gamma^{mm},
\end{equation}
and Eq.~(\ref{Eq:OCgeneral}) reduces to
\begin{equation}
(\mathcal J_\alpha^\gamma)^{mm'}
=
-\frac{1}{2}
\frac{\hbar}{g_L\mu_B}
\left(
\mathcal M_\gamma^{mm}
+
\mathcal M_\gamma^{m'm'}
\right)
v_\alpha^{mm'}.
\label{eq:orbital_current_matrix}
\end{equation}

For a nondegenerate Bloch band, the semiclassical orbital magnetic
moment is given by \cite{SongJustin2019}
\begin{equation}
\mathcal M_\gamma^{mm}
=
-\frac{e}{2\hbar}
\epsilon_{\gamma\lambda\eta}
\sum_{n\neq m}
(\varepsilon_m-\varepsilon_n)
\mathrm{Im}
\left[
\mathcal A_\lambda^{mn}
\mathcal A_\eta^{nm}
\right],
\label{eq:orbital_moment_definition}
\end{equation}
where \(\epsilon_{\gamma\lambda\eta}\) is the Levi-Civita symbol. Here and in what follows, the momentum dependence of the energies and matrix elements is left implicit whenever no ambiguity arises. 

For $m\neq m'$, the interband velocity matrix elements can be expressed in terms of the Berry connection as
\begin{equation}
v_\alpha^{mm'}
=
\frac{i}{\hbar}
(\varepsilon_m-\varepsilon_{m'})
\mathcal A_\alpha^{mm'}.
\label{eq:interband_velocity}
\end{equation}
Thus, the interband velocity is directly related to the geometrical
coupling between different Bloch bands, encoded in
$\mathcal A_\alpha^{mm'}$, with a weight determined by their energy
separation.

The orbital Hall conductivity is defined through the linear-response
relation
\begin{equation}
\langle\mathcal J_\alpha^\gamma\rangle
=
\sigma_{\alpha\beta}^{\gamma,\mathrm{Hall}}E_\beta.
\label{eq:OHC_definition}
\end{equation}
The intrinsic contribution to the expectation value of the orbital
current is obtained from the electric-field-induced interband density
matrix,
\begin{equation}
\langle\mathcal J_\alpha^\gamma\rangle
=
\sum_{\bm k}
\sum_{m\neq m'}
(\mathcal J_\alpha^\gamma)^{mm'}
\rho_E^{m'm}.
\label{eq:orbital_current_expectation}
\end{equation}
Using Eqs.~(\ref{Eq:offdiagonal}), (\ref{eq:orbital_current_matrix}),
and (\ref{eq:interband_velocity}), we obtain
\begin{align}
\langle\mathcal J_\alpha^\gamma\rangle
&=
\frac{ieE_\beta}{2g_L\mu_B}
\sum_{\bm k}
\sum_{m\neq m'}
\left(
\mathcal M_\gamma^{mm}
+
\mathcal M_\gamma^{m'm'}
\right)
\nonumber\\
&\qquad\times
\mathcal A_\alpha^{mm'}
\mathcal A_\beta^{m'm}
\left[
n_F(\varepsilon_{m'})
-
n_F(\varepsilon_m)
\right].
\label{eq:orbital_current_intermediate}
\end{align}
The terms related by the interchange $m\leftrightarrow m'$ can now be
combined. Using that $\mathcal A_\alpha^{m'm}=\left(\mathcal A_\alpha^{mm'}\right)^*$,
one finds that only the imaginary part of the product of interband
Berry connections contributes to the transverse response. The intrinsic
orbital Hall conductivity can therefore be written as
%
\begin{align}
\label{eq:sigma_general_hall}
\sigma_{\alpha\beta}^{\gamma,\mathrm{Hall}}
&=
\frac{e}{g_L\mu_B}
\sum_{\bm k}
\sum_{m\neq m'}
\left(
\mathcal M_\gamma^{mm}
+
\mathcal M_\gamma^{m'm'}
\right) \\ \nonumber
&\times 
\mathrm{Im}
\left[
\mathcal A_\alpha^{mm'}
\mathcal A_\beta^{m'm}
\right]
n_F(\varepsilon_m).
\end{align}
%
Equation~(\ref{eq:sigma_general_hall}) provides the general intrinsic
orbital Hall response within the present density-matrix formulation.
It makes explicit that the transverse orbital current originates from
the interplay between the orbital magnetic moment of the Bloch bands
and the interband geometry encoded in products of Berry connections.
In particular, the antisymmetric imaginary part
$\mathrm{Im}[
\mathcal A_\alpha^{mm'}
\mathcal A_\beta^{m'm}]$
is directly related to the Berry-curvature structure associated with
the pair of bands $m$ and $m'$. The specific form taken by this
geometrical contribution in the spin-$1$ model will be analyzed in the
following section.

\section{Results}

We now apply the theoretical framework developed above to investigate
orbital transport in the spin-$1$ model. We analyze the intrinsic
orbital Hall response and its dependence on the electronic structure
and Fermi level, identifying the contributions arising from the
different bands and interband channels.
The effects of temperature and the model parameters are also discussed.

\subsection{Intrinsic orbital Hall response of spin-1 fermions}

\subsubsection{Band-resolved orbital magnetic moment}

We begin by evaluating the orbital magnetic moment carried by each
helicity band. As follows from Eq.~\eqref{eq:orbital_moment_definition},
the orbital moment combines two distinct ingredients: the interband
geometry encoded in the Berry-connection matrix elements and the
energy separation between the participating bands.

For the spin-$1$ Hamiltonian introduced above, the eigenstates are
determined by the helicity operator $\hat{\bm k}\cdot\bm S$ and are
therefore independent of the quadratic parameter $c$. The latter
modifies the dispersion of the central band Eq.~(\ref{eq:spectrum}), 
without changing the helicity eigenstates or their Berry connections.
Consequently, $c$ enters the orbital magnetic moment through the
interband energy differences appearing in
Eq.~\eqref{eq:orbital_moment_definition}. This provides a useful
separation between the geometry of the eigenstates and the energetic
weighting of the corresponding interband processes.

The relevant Berry-connection products, derived in Appendix~\ref{app:berry_connection_products}, satisfy
%
\begin{align}
\operatorname{Im}
\left[
\mathcal A_\alpha^{0\pm}
\mathcal A_\beta^{\pm0}
\right]
&=
\pm\frac{1}{2}
\frac{\epsilon_{\alpha\beta\mu}k_\mu}{k^3},
\label{eq:berry_0pm_results}
\\ \nonumber
\operatorname{Im}
\left[
\mathcal A_\alpha^{+-}
\mathcal A_\beta^{-+}
\right]
&=0.
\end{align}
The second relation shows that the direct $+\leftrightarrow-$ channel
does not contribute to the orbital magnetic moment. Consequently, the
band-resolved orbital moments are entirely determined by virtual
transitions involving the central band.

We first consider the central band. Since it is geometrically coupled
to both dispersive bands, its orbital magnetic moment is
\begin{equation}
\mathcal M_\gamma^{00}
=
-\frac{e}{2\hbar}
\epsilon_{\gamma\alpha\beta}
\sum_{s=\pm 1}
(\varepsilon_0-\varepsilon_s)
\operatorname{Im}
\left[
\mathcal A_\alpha^{0s}
\mathcal A_\beta^{s0}
\right].
\label{eq:M0_start}
\end{equation}
Using
\begin{equation}
\varepsilon_0-\varepsilon_s
=
-ck^2-s\hbar v_F k
\end{equation}
together with Eq.~\eqref{eq:berry_0pm_results}, we obtain
\begin{align}
\mathcal M_\gamma^{00}
&=
-\frac{e}{2\hbar}
\epsilon_{\gamma\alpha\beta}
\sum_{s=\pm 1}
\left(-ck^2-s\hbar v_F k\right)
s\frac{\epsilon_{\alpha\beta\mu}k_\mu}{2k^3}.
\label{eq:M0_intermediate}
\end{align}
The contributions proportional to $c$ cancel between the two virtual
channels,
\begin{equation}
\sum_{s=\pm 1}s(-ck^2)=0,
\end{equation}
whereas the contributions associated with the linear pseudospin
coupling add,
\begin{equation}
\sum_{s=\pm 1}s(-s\hbar v_F k)
=
-2\hbar v_F k.
\end{equation}
Using $\epsilon_{\gamma\alpha\beta}\epsilon_{\alpha\beta\mu}=2\delta_{\gamma\mu}$
%
we therefore find
\begin{equation}
\mathcal M_\gamma^{00} = e v_F\frac{k_\gamma}{k^2}.
\label{eq:M0_final}
\end{equation}
Thus, although the central band has a quadratic dispersion, its
orbital magnetic moment is controlled entirely by the linear
pseudospin coupling. The quadratic contribution cancels because the
central band couples geometrically to both dispersive bands with
opposite geometric factors.

For the upper and lower dispersive bands, the direct
$+\leftrightarrow-$ channel gives no contribution according to
Eq.~\eqref{eq:berry_0pm_results}. Each dispersive band therefore
receives a contribution only from its virtual coupling to the central
band. Using Hermiticity of the Berry connection,
\begin{equation}
\mathcal A_\alpha^{m'm}
=
\left(
\mathcal A_\alpha^{mm'}
\right)^*,
\end{equation}
together with Eq.~\eqref{eq:berry_0pm_results}, one obtains
\begin{align}
\mathcal M_\gamma^{++}
&=
\frac{e}{2\hbar}
\left( \hbar v_F k+ck^2 \right)
\frac{k_\gamma}{k^3},
\label{eq:Mplus_final}
\\
\mathcal M_\gamma^{--}
&=
\frac{e}{2\hbar}
\left( \hbar v_F k-ck^2 \right)
\frac{k_\gamma}{k^3}.
\label{eq:Mminus_final}
\end{align}
Equivalently, the three band-resolved orbital magnetic moments can be
written in vector form as
\begin{equation}
\begin{aligned}
\boldsymbol{\mathcal M}_{0}(\bm k)
&=
e v_F\frac{\bm k}{k^2},
\\[3pt]
\boldsymbol{\mathcal M}_{+}(\bm k)
&=
\frac{e}{2}
\left(
v_F \pm \frac{c}{\hbar}k
\right)
\frac{\bm k}{k^2},
\end{aligned}
\label{eq:orbital_moments_summary}
\end{equation}
These expressions illustrate the different roles played by the
eigenstate geometry and the band dispersion. While $c$ leaves the
Berry connections unchanged, it modifies the orbital moments of the
two dispersive bands through their energy separation from the central
band. Its contribution enters with opposite signs in the $+$ and $-$
bands and cancels completely in the central-band orbital magnetic moment.

An immediate consequence of Eq.~\eqref{eq:orbital_moments_summary} is
the relation $\boldsymbol{\mathcal M}_{0}
=\boldsymbol{\mathcal M}_{+}+\boldsymbol{\mathcal M}_{-}$, which reflects the geometric coupling of the central band to both
dispersive bands and directly reveals the two-channel structure of
the spin-1 geometry. In the limit $c\rightarrow0$,
it reduces to $\boldsymbol{\mathcal M}_{+}=\boldsymbol{\mathcal M}_{-}=\frac{1}{2}\boldsymbol{\mathcal M}_{0}$, making explicit the factor of two associated with the two virtual
channels contributing to the central-band orbital magnetic moment.

\subsubsection{Intrinsic orbital Hall conductivity: general expression}

Having determined the band-resolved orbital magnetic moments, we now
evaluate the intrinsic orbital Hall conductivity using the general
expression derived in Eq.~\eqref{eq:sigma_general_hall}. For the
spin-$1$ eigenstates, the interband Berry connections obey the
selection rules associated with the Wigner--Eckart theorem. Since the
momentum derivatives of the helicity eigenstates generate infinitesimal
rotations and therefore involve rank-one angular-momentum operators,
interband matrix elements are nonzero only between adjacent helicity
states, $\Delta m=\pm1$. Consequently, the transitions
$0\leftrightarrow+$ and $0\leftrightarrow-$ are allowed, whereas the
direct $+\leftrightarrow-$ transition, for which $|\Delta m|=2$, is
forbidden \cite{BurgosAtencia2026}. Accordingly, $\operatorname{Im}\left[ \mathcal A_\alpha^{+-} \mathcal A_\beta^{-+} \right]=0$, as stated before.
The intrinsic orbital Hall response is therefore entirely determined
by the two interband channels connecting the central band with the
dispersive bands.

It is convenient to label the latter by $s=\pm 1$, such that
\begin{equation}
\varepsilon_s=s\hbar v_F k,
\qquad
\varepsilon_0=-ck^2.
\end{equation}
For each $0\leftrightarrow s$ channel, the relevant geometric factors
are given by Eq.~\eqref{eq:berry_0pm_results}. Combining the two
ordered contributions $(0,s)$ and $(s,0)$, the intrinsic orbital Hall
conductivity can be written as
\begin{align}
\sigma_{\alpha\beta}^{\gamma,\mathrm{Hall}}
&=
\frac{e}{g_L\mu_B}
\sum_{\bm k}
\sum_{s=\pm 1}
\left(
\mathcal M_\gamma^{00}
+
\mathcal M_\gamma^{ss}
\right)
s\,
\frac{\epsilon_{\alpha\beta\mu}k_\mu}{2k^3}
\\ \nonumber
&\times \left[
n_F(\varepsilon_0)
-
n_F(\varepsilon_s)
\right].
\label{eq:sigma_general_hall_spin1}
\end{align}
The response of each interband channel is therefore weighted by the
occupation difference between the two participating bands. In
particular, momentum regions in which the two bands have identical
occupations do not contribute to the intrinsic orbital Hall response.

Using the orbital moments in
Eq.~\eqref{eq:orbital_moments_summary}, the combination entering each
$0\leftrightarrow s$ channel becomes
\begin{equation}
\mathcal M_\gamma^{00}
+
\mathcal M_\gamma^{ss}
=
\frac{e}{2}
\left(
3v_F+s\frac{c}{\hbar}k
\right)
\frac{k_\gamma}{k^2}.
\label{eq:Msum_channel}
\end{equation}

The orbital Hall conductivity therefore takes the form
\begin{align}
\sigma_{\alpha\beta}^{\gamma,\mathrm{Hall}}
&=
\frac{e^2}{4g_L\mu_B}
\sum_{\bm k}
\sum_{s=\pm 1}
s
\left(
3v_F+s\frac{c}{\hbar}k
\right)
\frac{k_\gamma k_\mu}{k^5}
\epsilon_{\alpha\beta\mu}
\nonumber\\
&\qquad\times
\left[
n_F(\varepsilon_0)
-
n_F(\varepsilon_s)
\right].
\label{eq:OHC_before_angular}
\end{align}

For the isotropic model considered here, the angular dependence is
contained entirely in $k_\gamma k_\mu$. Using
\begin{equation}
\int d\Omega_{\bm k}\,
k_\gamma k_\mu
=
\frac{4\pi}{3}k^2\delta_{\gamma\mu},
\label{eq:angular_average}
\end{equation}
the tensorial structure of the response is fixed by rotational
invariance $\sigma_{\alpha\beta}^{\gamma,\mathrm{Hall}}=\epsilon_{\alpha\beta\gamma}\sigma_{\mathrm{OH}}$.
%
In the following we consider, without loss of generality,
$\alpha=x$, $\beta=y$, and $\gamma=z$, for which
$\sigma_{xy}^{z,\mathrm{Hall}}=\sigma_{\mathrm{OH}}$.
Replacing the momentum sum by
\begin{equation}
\sum_{\bm k}
\longrightarrow
\int\frac{d^3k}{(2\pi)^3},
\end{equation}
and performing the angular integration, we obtain
\begin{align}
\label{eq:OHC_radial}
\sigma_{xy}^{z,\mathrm{Hall}}
&=
\frac{e^2}{24\pi^2g_L\mu_B}
\sum_{s=\pm 1}
\int_0^\infty
\frac{dk}{k}\,
s
\left(
3v_F+s\frac{c}{\hbar}k
\right)
\\ \nonumber 
&\times \left[
n_F(\varepsilon_{0})
-
n_F(\varepsilon_{s})
\right]
\end{align}

Equation~\eqref{eq:OHC_radial} is the central expression for the
intrinsic orbital Hall conductivity of the spin-$1$ model. It shows
explicitly that the response is governed by the two interband channels
connecting the central band with the $s=\pm 1$ dispersive bands, while
their relative contributions are determined by the corresponding
occupation differences.

An important feature of Eq.~\eqref{eq:OHC_radial} appears at the
finite-momentum band crossing $k=k_c$, where the central band becomes
degenerate with one of the dispersive bands. For the corresponding
interband channel, $\varepsilon_0(k_c)=\varepsilon_s(k_c)$, and hence
the contribution of this channel to the orbital Hall
integrand vanishes exactly at the band crossing. This
cancellation does not originate from a vanishing geometric matrix
element: the $0\leftrightarrow s$ transition is allowed by the
spin-$1$ selection rules. Rather, it follows from the identical
occupation of the two degenerate states. The remaining
$0\leftrightarrow -s$ channel is not affected by this degeneracy and
can still contribute to the orbital Hall response.

It is convenient to retain the decomposition of the bare response into the contributions associated with the linear Dirac sector and the quadratic sector,
\begin{equation}
\sigma_{xy}^{z,\mathrm{Hall}}
=
\sigma_L+\sigma_Q,
\end{equation}
with
\begin{align}
\sigma_L=
\frac{3 e^2 v_F}{24\pi^2g_L\mu_B} 
\int_0^\infty\frac{dk}{k}\,
\left[
n_{F}(\varepsilon_{-})
-
n_{F}(\varepsilon_{+})
\right] 
\label{eq:sigma_L}
\end{align}
and
\begin{align}
\sigma_Q &=
\frac{e^2}{24\pi^2g_L\mu_B}
\left(\frac{c}{\hbar}\right)
\int_0^\infty dk \nonumber \\
&\times
\left[2 n_{F}(\varepsilon_{0}) -n_{F}(\varepsilon_{-})-n_{F}(\varepsilon_{+})\right].
\label{eq:sigma_Q}
\end{align}

The two terms have qualitatively different momentum dependence: the Dirac contribution contains the logarithmic measure $dk/k$, whereas the quadratic contribution is weighted by a constant radial measure.

\subsubsection{Intrinsic orbital Hall conductivity in the $T=0$ limit}

We first consider the Dirac sector, as discussed in Appendix~\ref{app:zero_temperature}. For the linear bands
$\varepsilon_\pm(k)=\pm\hbar v_F k$, the difference entering
$\sigma_L$ is particularly simple. Since at zero temperature
\begin{equation}
n_F(\varepsilon-\mu)
\underset{T\rightarrow0}{\longrightarrow}
\Theta(\mu-\varepsilon),
\end{equation}
we find
\begin{equation}
n_{F}(\varepsilon_{-})-n_{F}(\varepsilon_{+})
= \Theta\!\left(\mu+\hbar v_F k\right)-
\Theta\left(\mu-\hbar v_F k\right).
\label{eq:Delta_n_Dirac_T0}
\end{equation}

Introducing the magnitude of the Fermi momentum
$k_F=|\mu|/\hbar v_F$, the two possible signs of $\mu$ simply determine
which of the two Dirac bands crosses the Fermi level. For $\mu>0$, the
crossing occurs in the upper band, whereas for $\mu<0$ it occurs in the
lower band. In either case, the surviving occupation difference is
confined to the interval $k_F<k<\Lambda$. Hence,
\begin{equation}
n_{F}(\varepsilon_{-})-n_{F}(\varepsilon_{+})
\underset{T\rightarrow0}{\longrightarrow}
\Theta(k-k_F).
\label{eq:Delta_n_Dirac_piecewise}
\end{equation}

Consequently, the bare Dirac contribution at $T=0$ takes the logarithmic
form
\begin{equation}
\sigma_L
=
\frac{e^2v_F}{8\pi^2g_L\mu_B}
\ln\left(
\frac{\hbar v_F\Lambda}{|\mu|}
\right).
\label{eq:sigmaL_T0}
\end{equation}

Since the continuum Hamiltonian Eq.~\eqref{eq:H0} represents an effective
low-energy description of the band structure, it is not expected to remain
valid at arbitrarily large momenta. We therefore introduce an ultraviolet
momentum cutoff $\Lambda$, which sets the momentum scale beyond which the
effective model ceases to provide an accurate description of the
underlying band structure.

At zero temperature, the contribution associated with the quadratic band
follows from the same decomposition introduced above; the corresponding
derivation is given in Appendix~\ref{app:zero_temperature}.
Two momentum scales determine the changes in the occupation factors. The
first is the Dirac Fermi momentum $k_F$, which is present for either sign
of $\mu$. The second is the momentum at which the quadratic band crosses
the chemical potential,
\begin{equation}
k_0=\sqrt{\frac{|\mu|}{|c|}},
\end{equation}
which exists as a physical momentum only when $\mu c<0$. Thus, the
quadratic band contributes an additional momentum scale only when it
intersects the Fermi level.

Evaluating the piecewise-constant function
\begin{equation}
I_Q(k)
=
2n_{F}(\varepsilon_{0})
-n_{F}(\varepsilon_{-})
-n_{F}(\varepsilon_{+})
\end{equation}
gives a compact form for the quadratic contribution (See Appendix \ref{app:zero_temperature}),
\begin{equation}
\sigma_Q
=
\frac{e^2|c|}{24\pi^2\hbar g_L\mu_B}
\left[
\Lambda-k_F
+
2\Theta(-\mu c)
\left(k_F-k_0\right)
\right].
\label{eq:sigmaQ_compact}
\end{equation}

This expression contains both regimes in a single equation. For $\mu c>0$,
the quadratic band does not cross the Fermi level and
\begin{equation}
\sigma_Q
=
\frac{e^2|c|}{24\pi^2\hbar g_L\mu_B}
(\Lambda-k_F),
\end{equation}
whereas for $\mu c<0$ the additional crossing at $k_0$ modifies the result
according to
\begin{equation}
\sigma_Q
=
\frac{e^2|c|}{24\pi^2\hbar g_L\mu_B}
(\Lambda+k_F-2k_0).
\end{equation}

Combining the two sectors, the zero-temperature intrinsic orbital Hall
conductivity can therefore be written as
\begin{equation}
\begin{aligned}
\sigma_{xy}^{z,\mathrm{Hall}}(\mu,0)
= {}&
\frac{e^2v_F}{8\pi^2g_L\mu_B}
\ln\left(
\frac{\hbar v_F\Lambda}{|\mu|}
\right)  \\[-1mm]
+ {}&
\frac{e^2|c|}{24\pi^2\hbar g_L\mu_B}
\left[
\Lambda-k_F
+
2\Theta(-\mu c)
(k_F-k_0)
\right].
\end{aligned}
\label{eq:sigma_Hall_T0_total}
\end{equation}

This result displays two qualitatively different types of chemical-potential. The linear bands produce the logarithmic
dependence on $|\mu|$, whereas the quadratic band introduces a
piecewise contribution through $k_F\propto|\mu|$ and, when it crosses
the Fermi level, $k_0\propto\sqrt{|\mu|}$. At the same time, the
explicitly cutoff-dependent contributions are independent of the chemical
potential. These ultraviolet terms reflect the limited range of validity
of the continuum model.

To isolate the low-energy dependence on the chemical potential, we introduce a natural reference scale associated with the finite-momentum crossing between the quadratic band and the corresponding linearly dispersing branch. The
crossing occurs at
\begin{equation}
k_R=\frac{\hbar v_F}{|c|},
\qquad
\mu_R=\hbar v_F k_R
=\frac{(\hbar v_F)^2}{|c|},
\label{eq:reference_scales}
\end{equation}
where $\mu_R>0$ defines the characteristic energy scale associated with
this crossing. The corresponding chemical potential is
\begin{equation}
\mu_{\mathrm{ref}}
=
-\operatorname{sgn}(c)\mu_R.
\end{equation}
Thus, for $c>0$ the quadratic band crosses the negative-energy dispersive
branch at $\mu_{\mathrm{ref}}=-\mu_R$, whereas for $c<0$ the crossing
occurs at $\mu_{\mathrm{ref}}=+\mu_R$. At either crossing, the two
characteristic Fermi momenta coincide,
$k_F=k_0=k_R$. We therefore characterize the low-energy variation of the orbital Hall
response through the reference-subtracted conductivity
\begin{equation}
\Delta\sigma_{xy}^{z}(\mu)
=\sigma_{xy}^{z,\mathrm{Hall}}(\mu,0)-
\sigma_{xy}^{z,\mathrm{Hall}}(\mu_{\mathrm{ref}},0).
\label{eq:reference_subtracted_OHC}
\end{equation}

This subtraction removes the ultraviolet contribution independent of the chemical potential that cannot be determined within the low-energy continuum
theory.

For the linearly dispersing contribution, the subtraction gives
\begin{align}
\Delta\sigma_L(\mu)
&=
\frac{e^2v_F}{8\pi^2g_L\mu_B}
\ln\left(\frac{k_R}{k_F}\right).
\label{eq:sigmaL_subtracted}
\end{align}

The same procedure can be applied to the quadratic contribution. From
Eq.~\eqref{eq:sigmaQ_compact}, its value at the reference point is
\begin{equation}
\sigma_Q(\mu_{\mathrm{ref}})
=
\frac{e^2|c|}{24\pi^2\hbar g_L\mu_B}
(\Lambda-k_R),
\end{equation}
since $k_F=k_0=k_R$ at the band crossing. Consequently,
\begin{equation}
\Delta\sigma_Q(\mu)
=
\frac{e^2|c|}{24\pi^2\hbar g_L\mu_B}
\left[
k_R-k_F
+
2\Theta(-\mu c)
(k_F-k_0)
\right].
\label{eq:sigmaQ_subtracted}
\end{equation}

Equation~\eqref{eq:sigmaQ_subtracted} contains two physically distinct
regimes. When $\mu c>0$, the quadratic band does not intersect the Fermi
level, and
\begin{equation}
\Delta\sigma_Q(\mu)
=
\frac{e^2|c|}{24\pi^2\hbar g_L\mu_B}
(k_R-k_F).
\label{eq:sigmaQ_subtracted_same_sign}
\end{equation}

In contrast, when $\mu c<0$, the quadratic band crosses the Fermi level
and introduces the additional momentum scale $k_0$, yielding
\begin{equation}
\Delta\sigma_Q(\mu)
=
\frac{e^2|c|}{24\pi^2\hbar g_L\mu_B}
(k_R+k_F-2k_0).
\label{eq:sigmaQ_subtracted_opposite_sign}
\end{equation}

The appearance of $k_0\propto\sqrt{|\mu|}$ therefore provides a
characteristic signature of the quadratic dispersion in the Fermi-level
dependence of the orbital Hall response.

The zero-temperature result can be expressed in a dimensionless form
that makes the characteristic energy and momentum scales explicit.
Using the reference scales defined in Eq.~\eqref{eq:reference_scales}, we introduce the dimensionless chemical potential
\begin{equation}
x=\frac{\mu}{\mu_R}.
\end{equation}
The band crossing taken as the reference point is then located at
\begin{equation}
x_{\mathrm{ref}}
=
-\operatorname{sgn}(c),
\qquad
q_{\mathrm{ref}}=1,
\label{eq:dimensionless_reference_point}
\end{equation}
where $q=k/k_R$. 
Hence, $x_{\mathrm{ref}}=-1$ for $c>0$ and
$x_{\mathrm{ref}}=+1$ for $c<0$. In terms of these variables, the Fermi
momentum of the linearly dispersing branch becomes
$k_F=k_R|x|$, whereas, when $\mu c<0$, the quadratic-band Fermi momentum
is $k_0=k_R\sqrt{|x|}$. The two momentum scales therefore display
qualitatively different dependences on the chemical potential,
namely $k_F\propto|x|$ and $k_0\propto\sqrt{|x|}$.

For convenience, we normalize the reference-subtracted orbital Hall
conductivity according to
\begin{equation}
\Delta\bar{\sigma}_{xy}^{z}
\equiv
\frac{24\pi^2g_L\mu_B}{e^2v_F}
\Delta\sigma_{xy}^{z}.
\label{eq:dimensionless_conductivity}
\end{equation}

The contribution from the linearly dispersing bands then reduces to
\begin{equation}
\Delta\bar{\sigma}_L(x)
=
-3\ln|x|.
\label{eq:sigmaL_dimensionless_T0}
\end{equation}

This contribution is independent of the sign of $c$ and therefore
provides the same logarithmic dependence for the two possible
orientations of the quadratic band.

For the quadratic contribution, using
$|c|k_R/\hbar=v_F$, Eq.~\eqref{eq:sigmaQ_subtracted} becomes
\begin{equation}
\Delta\bar{\sigma}_Q(x)
=
1-|x|
+
2\Theta\!\left[-x\,\operatorname{sgn}(c)\right]
\left(
|x|-\sqrt{|x|}
\right).
\label{eq:sigmaQ_dimensionless_T0}
\end{equation}

Equivalently,
\begin{equation}
\Delta\bar{\sigma}_Q(x)
=
\begin{cases}
1-|x|,
&
x,\operatorname{sgn}(c)>0,
\\[2mm]
1+|x|-2\sqrt{|x|},
&
x,\operatorname{sgn}(c)<0.
\end{cases}
\label{eq:sigmaQ_dimensionless_piecewise}
\end{equation}

The first regime corresponds to a Fermi level that does not intersect
the quadratic band. In the second regime, the quadratic band participates
at the Fermi level and the additional scale
$k_0\propto\sqrt{|x|}$ becomes active.

Combining the two contributions, the complete reference-subtracted
orbital Hall conductivity at zero temperature is
\begin{equation}
\begin{aligned}
\Delta\bar{\sigma}_{xy}^{z}(x,0)
&=
-3\ln|x|+1-|x|
\\
&\quad
+
2\Theta\!\left[-x\,\operatorname{sgn}(c)\right]
\left(
|x|-\sqrt{|x|}
\right).
\end{aligned}
\label{eq:sigma_total_dimensionless_T0}
\end{equation}

or, equivalently,
\begin{equation}
\Delta\bar{\sigma}_{xy}^{z}(x,0)
=
\begin{cases}
-3\ln|x|+1-|x|,
&
x\,\operatorname{sgn}(c)>0,
\\[2mm]
-3\ln|x|+1+|x|-2\sqrt{|x|},
&
x\,\operatorname{sgn}(c)<0.
\end{cases}
\label{eq:sigma_total_dimensionless_piecewise}
\end{equation}

Several features of Eq.~\eqref{eq:sigma_total_dimensionless_T0} are
worth emphasizing. First, the contribution from the linearly dispersing
bands is insensitive to the orientation of the quadratic band and
produces a logarithmic dependence on the distance from the threefold
node. Second, the quadratic contribution distinguishes the two sides
of the spectrum through the condition
$x\,\operatorname{sgn}(c)<0$, where the quadratic band intersects the
Fermi level and generates the characteristic square-root dependence.

Finally, at the reference crossing
$x=x_{\mathrm{ref}}=-\operatorname{sgn}(c)$ one has
$|x_{\mathrm{ref}}|=1$ and therefore
\begin{equation}
\Delta\bar{\sigma}_L(x_{\mathrm{ref}})
=
\Delta\bar{\sigma}_Q(x_{\mathrm{ref}})
=
\Delta\bar{\sigma}_{xy}^{z}(x_{\mathrm{ref}},0)
=
0,
\end{equation}
as required by the definition of the reference-subtracted response.

To further characterize the dependence of the orbital Hall response on the
Fermi level, it is useful to consider its differential response with respect
to the chemical potential. This quantity directly measures how the orbital
Hall conductivity changes as the Fermi level is varied and provides a
convenient way to identify the features associated with the different
dispersive sectors of the spectrum. We therefore define the differential
orbital Hall response as
\begin{equation}
D_{xy}^{z}(\mu,0)
=
\frac{\partial \sigma_{xy}^{z,\mathrm{Hall}}(\mu,0)}
{\partial \mu}.
\end{equation}
An additional advantage of this quantity is that any contribution independent
of the chemical potential is automatically removed. In particular, since the
reference point $x_{\mathrm{ref}}=-\operatorname{sgn}(c)$ is fixed, the
reference subtraction contributes only a $\mu$-independent constant.
Consequently,
\begin{equation}
D_{xy}^{z}(\mu,0)
=
\frac{\partial \Delta\sigma_{xy}^{z}(\mu,0)}
{\partial\mu}.
\end{equation}

The differential Hall response follows directly from the dimensionless
form above. Since
\begin{equation}
\frac{\partial}{\partial\mu}
=
\frac{1}{\mu_R}
\frac{\partial}{\partial x},
\end{equation}
we obtain
\begin{equation}
D_{xy}^{z}(\mu,0)
=
\frac{e^2v_F}
{24\pi^2g_L\mu_B\mu_R}
\frac{\partial
\Delta\bar{\sigma}_{xy}^{z}(x,0)}
{\partial x}.
\label{eq:D_dimensionless_relation}
\end{equation}
For $x\neq0$, differentiation of
Eq.~\eqref{eq:sigma_total_dimensionless_T0} gives
\begin{equation}
\begin{aligned}
\bar{D}_{xy}^{z}(x,0)
\equiv{}&
\frac{24\pi^2g_L\mu_B\mu_R}{e^2v_F}
D_{xy}^{z}(\mu,0)\\
={}&
-\frac{3}{x}
-\operatorname{sgn}(x)
\\[-1mm]
&\quad
+2\Theta\!\left[-x\,\operatorname{sgn}(c)\right]
\operatorname{sgn}(x)
\left(
1-\frac{1}{2\sqrt{|x|}}
\right).
\end{aligned}
\label{eq:D_dimensionless_T0}
\end{equation}

The cutoff $\Lambda$ is absent from
Eq.~\eqref{eq:D_dimensionless_T0} because the explicitly ultraviolet
contribution to the zero-temperature conductivity is independent of
$\mu$. The differential response thus directly probes the chemical-potential dependence of the low-energy continuum model. In particular, the differential orbital Hall response separates the logarithmic Dirac contribution, which gives the characteristic $1/\mu$ dependence, from the curvature-induced contribution of the quadratic band, which gives an additional $1/\sqrt{|\mu|}$ singularity on the side $x,\operatorname{sgn}(c)<0$, where it intersects the Fermi level. At $x=0$, the continuum expression is singular, reflecting the nonanalyticity of the zero-temperature response at the threefold node.

\begin{figure}[t]
\centering
\includegraphics[width=\columnwidth]{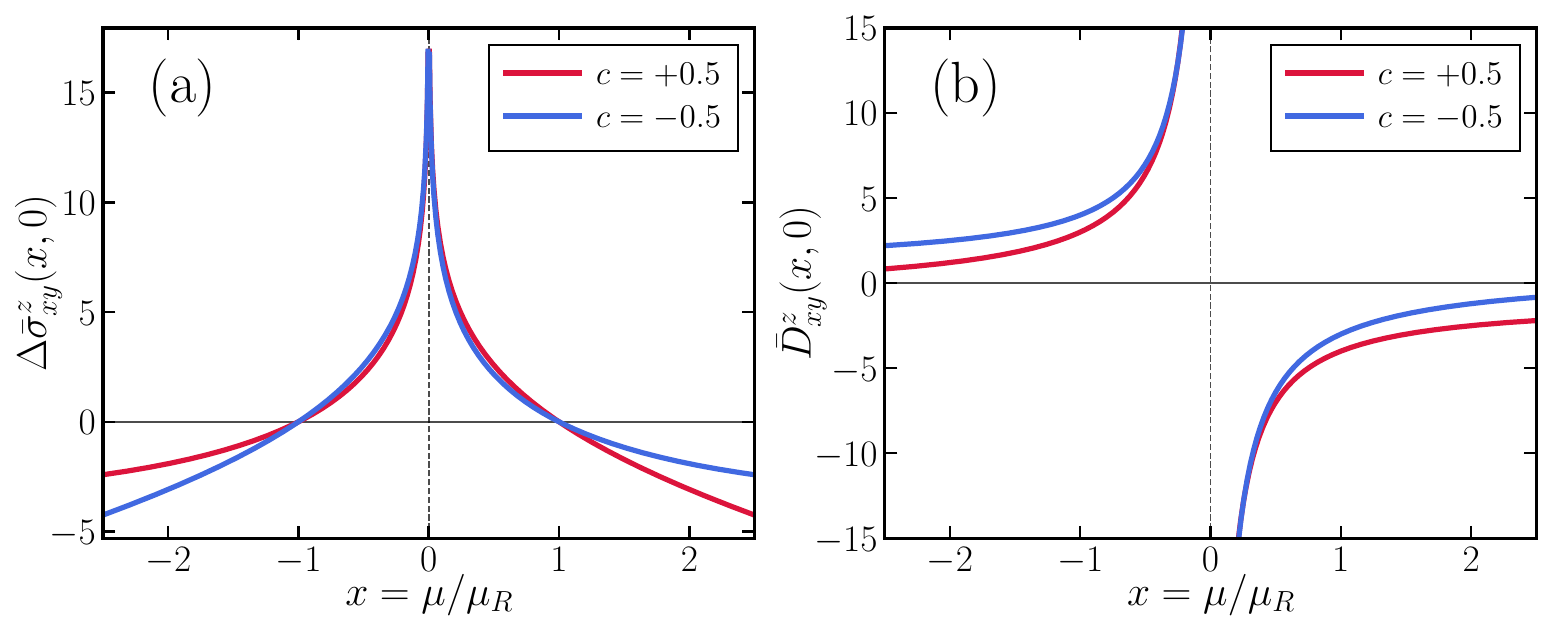}
\caption{Dimensionless Hall response and differential at $T=0$ as
functions of $x=\mu/\mu_{\mathrm R}$ for $c>0$ and $c<0$.
The responses are independent of $|c|$ and depend only on
$\operatorname{sgn}(c)$.}
\label{fig:hall_dimensionless_comparison}
\end{figure}

As shown in Fig.~\ref{fig:hall_dimensionless_comparison}, changing the sign of $c$ changes the side of the chemical-potential axis on which the contribution associated with the quadratic band becomes active. Thus, the effect of $c$ is not restricted to an overall rescaling of the response: its sign changes the structure of the Hall response for electron- and hole-like chemical potentials. This asymmetry follows directly from the factor $\operatorname{sgn}(c)$ entering the dimensionless response.

A pronounced feature appears around $x=0$, where both the Hall response and its derivative vary rapidly. This behavior reflects the low-energy structure of the model and the nonanalytic dependence of the response on the chemical potential near the band-crossing point. In particular, the differential response provides a more sensitive probe of this region, since it measures directly how rapidly the Hall response changes as the chemical potential is swept through the nodal energy. The sharp feature should therefore be understood as an intrinsic consequence of the low-energy band structure rather than as a numerical artifact.

In summary, although the Hall conductivity itself contains cutoff-dependent contributions, the terms responsible for this dependence are independent of the chemical potential and consequently vanish upon taking $\partial\sigma_{xy}^{z}/\partial\mu$. The resulting differential response therefore probes the chemical-potential-dependent part of the Hall conductivity without retaining the corresponding ultraviolet ambiguity. This also explains why the zero-temperature expression for the dimensionless differential response contains no explicit $\Lambda$.


\subsubsection{Intrinsic orbital Hall conductivity at finite temperature}

To extend the zero-temperature results to finite temperature, we introduce the dimensionless chemical potential, momentum, and temperature
\begin{equation}
x=\frac{\mu}{\mu_R},
\qquad
q=\frac{k}{k_R},
\qquad
t=\frac{k_BT}{\mu_R},
\label{eq:dimensionless_variables}
\end{equation}
where $t$ denotes the reduced temperature and should not be confused with time. In terms of these variables, the three band dispersions are
\begin{equation}
\frac{\varepsilon_\pm(k)}{\mu_R}=\pm q,
\qquad
\frac{\varepsilon_0(k)}{\mu_R}
=
-\operatorname{sgn}(c)q^2 .
\label{eq:dimensionless_bands}
\end{equation}

The corresponding occupation functions are
\begin{equation}
\begin{aligned}
n_\pm(q;x,t)
&=
\frac{1}{\exp[(\pm q-x)/t]+1},
\\
n_0(q;x,t)
&=
\frac{1}
{\exp\left[
\left(-\operatorname{sgn}(c)q^2-x\right)/t
\right]+1}.
\end{aligned}
\label{eq:dimensionless_occupations}
\end{equation}

We use the same occupation differences as in the dimensional formulation,
\begin{equation}
\Delta n_\pm(q;x,t)
=
n_0(q;x,t)-n_\pm(q;x,t),
\label{eq:dimensionless_delta_n}
\end{equation}
and define the two combinations associated with the Dirac and quadratic
sectors,
\begin{equation}
\begin{aligned}
\mathcal P(q;x,t)
&=
n_-(q;x,t)-n_+(q;x,t),
\end{aligned}
\label{eq:dimensionless_D}
\end{equation}
and
\begin{equation}
\begin{aligned}
\mathcal Q(q;x,t)
&=
2n_0(q;x,t)-n_+(q;x,t)-n_-(q;x,t).
\end{aligned}
\label{eq:dimensionless_Q}
\end{equation}

With the conductivity normalized as in
Eq.~\eqref{eq:dimensionless_conductivity}, the finite-temperature bare
response takes the form
\begin{equation}
\begin{aligned}
\bar{\sigma}_{xy}^{z,\mathrm{Hall}}(x,t)
= {}&
3\int_0^{\Lambda/k_R}\frac{dq}{q}\,
\mathcal P(q;x,t)
\\
&+
\operatorname{sgn}(c)
\int_0^{\Lambda/k_R}dq\,
\mathcal Q(q;x,t).
\end{aligned}
\label{eq:dimensionless_bare_response}
\end{equation}

The first integral retains the logarithmic measure characteristic of the Dirac contribution, whereas the second carries the constant radial measure associated with the quadratic dispersion. After Eq.~\eqref{eq:dimensionless_bare_response}, the dependence on the magnitude $|c|$ drops out, leaving only the dependence on the sign of $c$. Thus, this formulation preserves the decomposition established in the zero-temperature analysis.

The same reference-subtraction procedure is applied at fixed temperature. Since the matching chemical potential is
\begin{equation}
\mu_{\mathrm{ref}}
=
-\operatorname{sgn}(c)\mu_R,
\qquad
x_{\mathrm{ref}}=-\operatorname{sgn}(c),
\label{eq:finiteT_matching_point}
\end{equation}
the reference-subtracted response is defined by
\begin{equation}
\Delta\bar{\sigma}_{xy}^{z}(x,t)
=
\bar{\sigma}_{xy}^{z,\mathrm{Hall}}(x,t)
-
\bar{\sigma}_{xy}^{z,\mathrm{Hall}}(x_{\mathrm{ref}},t).
\label{eq:dimensionless_renormalized_response}
\end{equation}

Because the two terms are evaluated within the same continuum theory and at the same temperature, the chemical-potential-independent ultraviolet contribution cancels in the difference. The upper limit can therefore be extended to infinity, yielding the
convergent expression
\begin{equation}
\begin{aligned}
\Delta\bar{\sigma}_{xy}^{z}(x,t)
={}&
3\int_0^\infty\frac{dq}{q}
\left[
\mathcal P(q;x,t)
-
\mathcal P(q;x_{\mathrm{ref}},t)
\right]
\\
&+
\operatorname{sgn}(c)
\int_0^\infty dq
\left[
\mathcal Q(q;x,t)
-
\mathcal Q(q;x_{\mathrm{ref}},t)
\right].
\end{aligned}
\label{eq:dimensionless_renormalized_integral}
\end{equation}

This is the central expression used in the numerical calculations. Once
the conductivity is normalized by
$e^2v_F/(24\pi^2g_L\mu_B)$, the finite-temperature response is governed
by the reduced chemical potential $x$ and temperature $t$, with the sign
of $c$ distinguishing the two band geometries. In the limit
$t\rightarrow0$, Eq.~\eqref{eq:dimensionless_renormalized_integral}
continuously recovers the analytical results obtained above.

The differential Hall response is obtained from the same dimensionless
formulation used above. Since the reference point $x_{\mathrm{ref}}$ is
fixed, the reference subtraction does not contribute to the derivative
with respect to $x$. Thus,
\begin{equation}
D_{xy}^{z}(\mu,T)
=
\frac{e^2v_F}
{24\pi^2g_L\mu_B\mu_R}
\frac{\partial
\Delta\bar{\sigma}_{xy}^{z}(x,t)}
{\partial x}.
\label{eq:D_dimensionless_finiteT}
\end{equation}

Rather than differentiating the numerically evaluated conductivity, the
derivative can be taken analytically at the level of the Fermi functions.
For the occupations introduced above,
\begin{equation}
\frac{\partial n_a}{\partial x}
=
\frac{1}{t}\,n_a(1-n_a),
\qquad
a=+1, 0, -1.
\end{equation}
so that the dimensionless differential response becomes
\begin{equation}
\begin{aligned}
\bar{D}_{xy}^{z}(x,t)
\equiv{}&
\frac{24\pi^2g_L\mu_B\mu_R}
{e^2v_F}
D_{xy}^{z}(\mu,T)\\
={}&
\frac{1}{t}
\sum_{s=\pm 1}
\int_0^\infty\frac{dq}{q}\,
s\left[3+s\,\operatorname{sgn}(c)\,q\right]
\\
&\times
\left[
n_0(1-n_0)
-
n_s(1-n_s)
\right].
\end{aligned}
\label{eq:D_dimensionless_numerical}
\end{equation}

This form is particularly convenient for the numerical analysis because
the differential response is evaluated directly from the occupation
functions, without numerical differentiation of the Hall conductivity.
It involves only the same dimensionless variables and band occupations
introduced above, while the fixed reference contribution drops out
identically. In the zero-temperature limit, it continuously reduces to
Eq.~\eqref{eq:D_dimensionless_T0}.

\begin{figure}[t]
\centering
\includegraphics[width=\columnwidth]{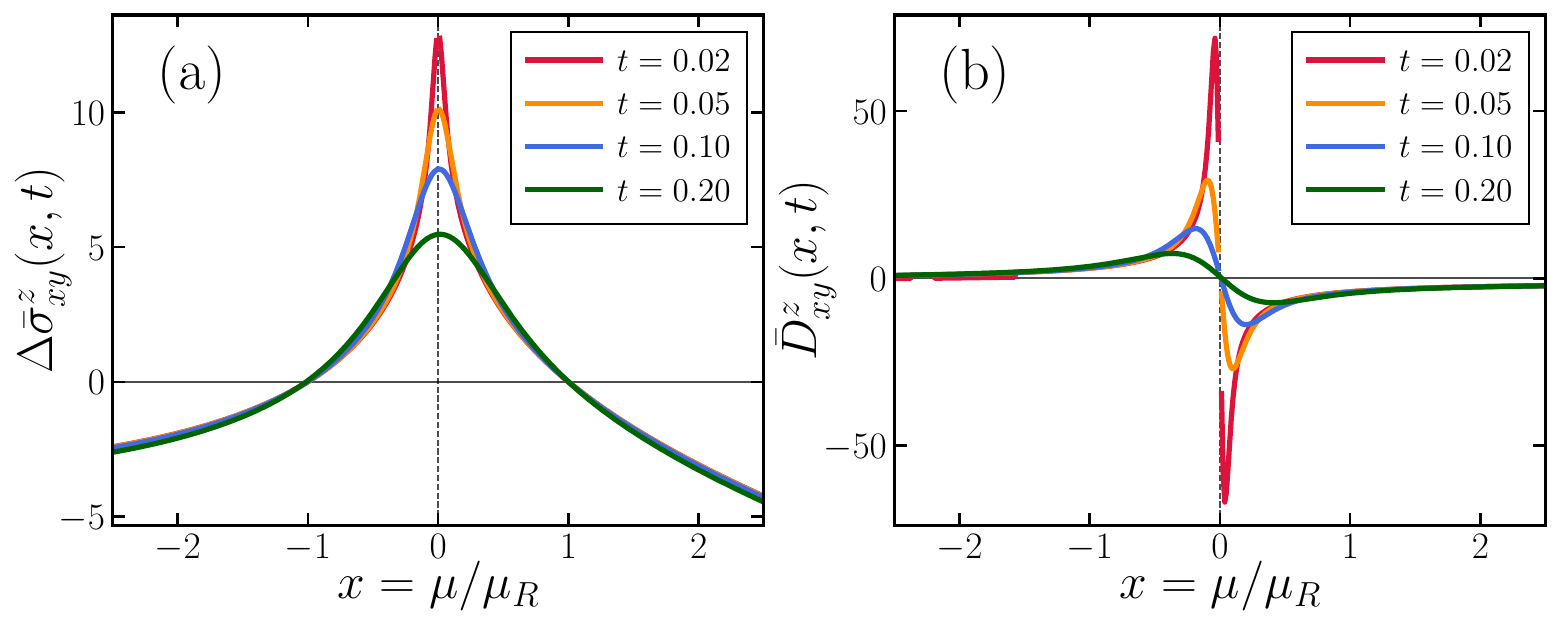}
\caption{Dimensionless Hall response at finite temperature for
$c=+0.5$. (a) Reference-subtracted Hall response
$\Delta\bar{\sigma}_{xy}^{z}(x,t)$; (b) differential Hall response
$\bar D_{xy}^{z}(x,t)$, for $t=0.02$, $0.05$, $0.10$, and $0.20$,
with $x=\mu/\mu_{\mathrm R}$.}
\label{fig:hall_temperature_dimensionless}
\end{figure}

The finite-temperature Hall response and its differential are shown in Fig.~\ref{fig:hall_temperature_dimensionless}. In this case, the sign of the quadratic parameter is fixed to $c>0$, since the effect of changing the sign of $c$ has already been established at zero temperature. The calculation is performed for the dimensionless temperatures $t=0.02$, $0.05$, $0.10$, and $0.20$, with the chemical potential expressed as $x=\mu/\mu_{\mathrm R}$.


As shown in Fig.~\ref{fig:hall_temperature_dimensionless}(a), increasing the temperature progressively smooths the dependence of the Hall response on the chemical potential. At the lowest temperature, the finite-temperature result closely approaches the zero-temperature behavior, including the rapid variation near $x=0$. With increasing $t$, the Fermi functions broaden the occupation of the bands over an energy window of order $t$, reducing the sharpness of the low-energy structure.

This thermal smoothing is even more evident in the differential response shown in Fig.~\ref{fig:hall_temperature_dimensionless}(b). The pronounced peak around $x=0$ present in the low-temperature limit becomes progressively broader and less sharp as the temperature increases. This behavior has a direct physical interpretation: $\bar D_{xy}^{z}$ measures the sensitivity of the Hall response to changes in the chemical potential, and thermal broadening averages this response over a finite energy window. Consequently, the singular or sharply varying zero-temperature structure is replaced by a finite-width thermal feature.

The comparison between the two panels also shows that the differential response is particularly sensitive to thermal effects. While the Hall conductivity retains the overall structure of the zero-temperature response over a relatively broad range of $x$, its derivative directly resolves the temperature-induced rounding of the low-energy feature. Thus, the differential response provides a convenient way to characterize the crossover from the sharp zero-temperature regime to the thermally broadened regime.

As in the zero-temperature case, the differential Hall response is independent of the ultraviolet cutoff $\Lambda$. In the renormalized formulation, $\Lambda$ does not appear explicitly in $\bar D_{xy}^{z}(x,t)$. More generally, any cutoff-dependent contribution to the Hall conductivity that is independent of the chemical potential is eliminated by the derivative with respect to $\mu$. Therefore, the temperature dependence displayed in Fig.~\ref{fig:hall_temperature_dimensionless}(b) represents the finite-energy occupation effects rather than an ultraviolet regularization effect.

We now analyze the asymptotic scaling of the finite-temperature
differential Hall response in the vicinity of the band-crossing point.
We focus on the regime 
\begin{equation}
    |\mu| \sim k_B T \ll \mu_R,
    \qquad
    |x|\sim t\ll1,
    \qquad
    X=\frac{x}{t}=O(1),
\end{equation}
where the chemical potential and temperature approach the crossing
simultaneously while their ratio remains finite. In this regime, the
linear and quadratic bands are thermally sampled over different
momentum scales, $q_T^{\rm L}\sim t$ and $q_T^{\rm Q}\sim\sqrt{t}$,
respectively. This difference generates distinct asymptotic powers of temperature in the differential Hall response.

Starting from the dimensionless expression in
Eq.~\eqref{eq:D_dimensionless_numerical}, the exact decomposition derived in Appendix~\ref{app:scaling_D} gives
\begin{equation}
    \bar D_{xy}^{z}(x,t)
    =
    \frac{1}{t}\mathcal{F}(X)
    +
    \frac{1}{\sqrt{t}}\mathcal{G}(X)
    +
    \mathcal{H}(X),
    \qquad X=\frac{x}{t}.
    \label{eq:D_scaling_complete}
\end{equation}
The three scaling functions are
\begin{align}
    \mathcal{F}(X)
    &=
    -3\int_0^\infty \frac{dy}{y}
    \left[
        h(y-X)-h(-y-X)
    \right],
    \label{eq:F_scaling}
    \\
    \mathcal{G}(X)
    &=
    2s_c\int_0^\infty dy\,
    h(-s_c y^2-X),
    \label{eq:G_scaling}
    \\
    \mathcal{H}(X)
    &=
    -s_c\int_0^\infty dy\,
    \left[
        h(y-X)+h(-y-X)
    \right],
    \label{eq:H_scaling}
\end{align}
where
\begin{equation}
    h(z)=\frac{e^z}{(1+e^z)^2}
    =\frac{1}{4\cosh^2(z/2)},
    \qquad
    s_c=\operatorname{sgn}(c).
\end{equation}

The derivation of Eq.~\eqref{eq:D_scaling_complete}, including the
separation of the linear and quadratic contributions and the associated
thermal momentum rescalings, is given in Appendix~\ref{app:scaling_D}. 
Equation Eq.~\eqref{eq:D_scaling_complete} provides an exact decomposition of the finite-temperature response. Its scaling structure emerges in the limit \(t\to0\) at fixed \(X=x/t\). 
In this limit, the three contributions are ordered as $t^{-1}\gg t^{-1/2}\gg t^0$, yielding
\begin{equation}
    \bar D_{xy}^{z}
    \sim
    \frac{\mathcal{F}(X)}{t}
    +
    \frac{\mathcal{G}(X)}{\sqrt{t}}
    + O(t^0),
    \label{eq:D_scaling_leading}
\end{equation}
which directly reflects the different thermal momentum scales of the
linear and quadratic sectors. The $t^{-1}$ contribution originates from
the Dirac-like bands, whose thermally active momenta scale as
$q_T^{\rm L}\sim t$, whereas the $t^{-1/2}$ contribution is generated
by the curved central band, for which
$q_T^{\rm Q} \sim \sqrt{t}$. Thus, the finite-temperature response
retains a direct signature of the different dispersions near the
crossing.

\begin{figure}[t]
\centering
\includegraphics[width=\linewidth]{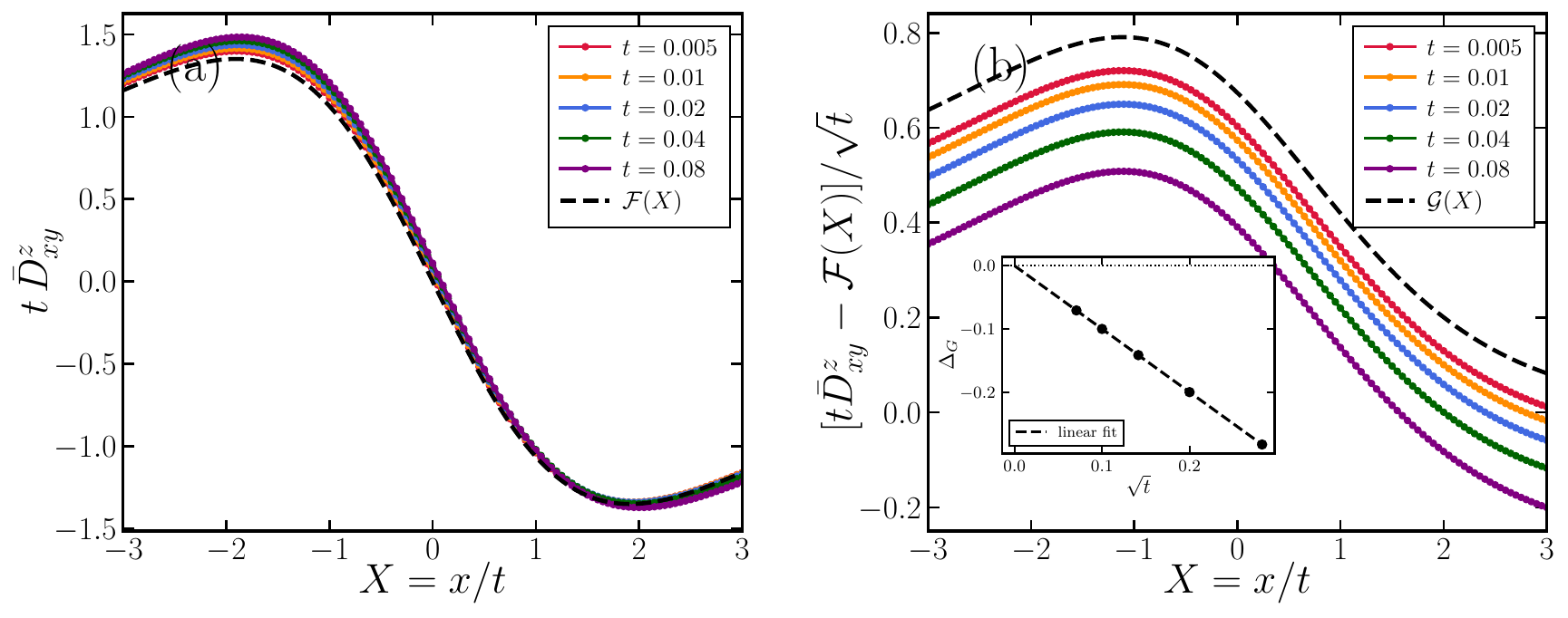}
\caption{Asymptotic scaling of the differential Hall response for
$|x|\sim t\ll1$. (a) $t\bar D_{xy}^{z}$ versus $X=x/t$ and the
leading scaling function $\mathcal{F}(X)$. (b)
$(t\bar D_{xy}^{z}-\mathcal{F})/\sqrt{t}$ versus $X$ and the
subleading function $\mathcal{G}(X)$. Deviations at finite $t$ are
governed by the $\sqrt{t}\,\mathcal{H}(X)$ correction. The inset
shows $\Delta_G(X_0,t)$ versus $\sqrt{t}$.}
\label{fig:scaling}
\end{figure}

A convenient way to test the leading scaling is to multiply
Eq.~\eqref{eq:D_scaling_complete} by $t$, yielding
\begin{equation}
    t\bar D_{xy}^{z}
    =
    \mathcal{F}(X)
    +
    \sqrt{t}\,\mathcal{G}(X)
    +
    t\,\mathcal{H}(X).
    \label{eq:tD_scaling}
\end{equation}
Consequently, at sufficiently low temperature, curves obtained at
different $t$ should approach the same function $\mathcal{F}(X)$ when
plotted as a function of $X=x/t$. This leading data collapse is shown
in Fig.~\ref{fig:scaling}(a). The deviations from the limiting curve
are controlled first by the curvature-induced contribution
$\sqrt{t}\,\mathcal{G}(X)$.

The subleading scaling can be isolated by subtracting the leading
contribution and dividing by $\sqrt{t}$:
\begin{equation}
    \frac{
        t\bar D_{xy}^{z}-\mathcal{F}(X)
    }{\sqrt{t}}
    =
    \mathcal{G}(X)
    +
    \sqrt{t}\,\mathcal{H}(X).
    \label{eq:subleading_scaling}
\end{equation}
Hence, this quantity approaches the scaling function
$\mathcal{G}(X)$ in the limit $t\rightarrow0$, while the remaining
finite-temperature deviation is predicted to be linear in $\sqrt{t}$.
The corresponding numerical test is presented in
Fig.~\ref{fig:scaling}(b).

For a fixed value $X=X_0$, it is useful to define
\begin{equation}
    \Delta_G(X_0,t)
    =
    \frac{
        t\bar D_{xy}^{z}(X_0,t)-\mathcal{F}(X_0)
    }{\sqrt{t}}
    -\mathcal{G}(X_0).
    \label{eq:Delta_G}
\end{equation}
Within the scaling decomposition above,
\begin{equation}
    \Delta_G(X_0,t)
    =
    \sqrt{t}\,\mathcal{H}(X_0).
    \label{eq:Delta_G_scaling}
\end{equation}
Thus, the appropriate test of the subleading scaling is the linear
dependence of $\Delta_G(X_0,t)$ on $\sqrt{t}$. The coefficient of this
linear dependence is $\mathcal{H}(X_0)$ and therefore depends on the
chosen value of $X_0$: it is not a universal coefficient.


Together, Eqs.~\eqref{eq:tD_scaling} and
\eqref{eq:subleading_scaling} provide two complementary tests of the
finite-temperature scaling. The first establishes the leading
collapse associated with the Dirac sector, while the second isolates
the curvature-induced correction and verifies its predicted
$t^{-1/2}$ contribution. The numerical results in
Fig.~\ref{fig:scaling} are consistent with this hierarchy and show that the deviations from the leading collapse are systematically controlled by
the subleading scaling structure.

\section{Conclusion}

We have investigated the intrinsic orbital Hall response of spin-1 fermions with a curved central band. Using a quantum Liouville formulation, we derive the orbital Hall conductivity in terms of interband coherence and orbital magnetic moments. The spin-1 selection rules restrict the response to the two channels connecting the central band with the dispersive bands. The quadratic parameter $c$ leaves the eigenstate geometry unchanged while modifying the band energies and orbital magnetic moments, thereby controlling the Hall response. This separates the contribution of the linearly dispersing bands from that of the curved central band and reveals the role of the curvature orientation in the response.
Because the model is a low-energy continuum theory, the bare orbital Hall conductivity depends on the ultraviolet cutoff. We remove this dependence by defining the response relative to the finite-momentum crossing between the central and dispersive bands. The resulting reference-subtracted conductivity is cutoff independent and directly captures its chemical-potential and temperature dependence. At zero temperature, the differential orbital Hall response separates the logarithmic Dirac contribution, with its characteristic $1/\mu$ dependence, from the curvature-induced square-root singularity, $\sim1/\sqrt{|\mu|}$, that appears when the central band crosses the Fermi level. At finite temperature, thermal broadening smooths these structures while preserving their distinct origins. These results establish central-band curvature as a direct control of orbital Hall transport in spin-1 fermions.

\appendix

\section{Berry-Connection Products for the Spin-1 Helicity States}
\label{app:berry_connection_products}

We derive the Berry-connection products entering Eq.~\eqref{eq:berry_0pm_results}. The helicity eigenstates are written as
\begin{equation}
|u_m(\theta,\phi)\rangle
=
U(\theta,\phi)|m\rangle_z,
\qquad
m=+1,0,-1,
\label{eq:app_rotstates}
\end{equation}
with the rotation operator 
\begin{equation}
U(\theta,\phi)
=
e^{-i\phi S_z}e^{-i\theta S_y}.
\label{eq:app_rotationoperator}
\end{equation}

The angles are defined through the relation 
$\hat{\bm k}=k(\sin\theta\cos\phi,\sin\theta\sin\phi,\cos\theta)$.

The Berry-connection matrix is
\begin{equation}
\mathcal A_{\alpha}^{mm'}
=
i\left\langle u_m\left|
\frac{\partial}{\partial k_\alpha}
\right|u_{m'}\right\rangle.
\label{eq:app_defberry}
\end{equation}

Equivalently in terms of the rotation operator it reads
\begin{equation}
\mathcal A=iU^\dagger dU.
\label{eq:app_berry_oneform}
\end{equation}

Using Eq.~\eqref{eq:app_rotationoperator},

\begin{align}
\mathcal A&=
e^{i\theta S_y}S_z e^{-i\theta S_y}\,d\phi
+
S_y\,d\theta .
\end{align}
The standard rotation identity
\begin{equation}
e^{i\theta S_y}S_z e^{-i\theta S_y}
=
S_z\cos\theta-S_x\sin\theta
\end{equation}
then gives
\begin{equation}
\boxed{
\mathcal A
=
S_y\,d\theta
+
\left(S_z\cos\theta-S_x\sin\theta\right)d\phi .
}
\label{eq:app_berry_connection}
\end{equation}

We now evaluate the matrix elements relevant for the products
$\mathcal A^{0\pm}\mathcal A^{\pm0}$.
In the $S_z$ basis, the spin-1 matrices are
\begin{equation}
\begin{aligned}
S_x &=
\frac{1}{\sqrt{2}}
\begin{pmatrix}
0&1&0\\
1&0&1\\
0&1&0
\end{pmatrix},
\\[4pt]
S_y &=
\frac{1}{\sqrt{2}}
\begin{pmatrix}
0&-i&0\\
i&0&-i\\
0&i&0
\end{pmatrix},
\\[4pt]
S_z &=
\begin{pmatrix}
1&0&0\\
0&0&0\\
0&0&-1
\end{pmatrix}.
\end{aligned}
\label{eq:app_spin1_matrices}
\end{equation}
Therefore,
\begin{align}
\langle0|S_x|\pm1\rangle
&=\frac{1}{\sqrt2},
&
\langle0|S_y|\pm1\rangle
&=\pm\frac{i}{\sqrt2},
\end{align}
where the upper sign corresponds to $m=+1$. Since $\langle0|S_z|\pm1\rangle=0$, Eq.~\eqref{eq:app_berry_connection} gives
\begin{align}
\mathcal A^{0\pm}
&=
\mp\frac{1}{\sqrt2}\,d\theta
-\frac{i}{\sqrt2}\sin\theta\,d\phi .
\label{eq:app_A0pm}
\end{align}

Since $\mathcal A=iU^\dagger dU$ is Hermitian,
\begin{equation}
\mathcal A^{\pm0}
=\left(\mathcal A^{0\pm}\right)^*=
\mp\frac{1}{\sqrt2}\,d\theta
+\frac{i}{\sqrt2}\sin\theta\,d\phi .
\label{eq:app_Apm0}
\end{equation}

Writing
\begin{equation}
\theta_\alpha\equiv\partial_{k_\alpha}\theta,
\qquad
\phi_\alpha\equiv\partial_{k_\alpha}\phi,
\end{equation}
we obtain
\begin{align}
\mathcal A_\alpha^{0\pm}
&=
\mp\frac{\theta_\alpha}{\sqrt2}
-\frac{i\sin\theta}{\sqrt2}\phi_\alpha,
\\ \nonumber
\mathcal A_\beta^{\pm0}
&=
\mp\frac{\theta_\beta}{\sqrt2}
+\frac{i\sin\theta}{\sqrt2}\phi_\beta.
\end{align}
Their product is therefore
\begin{align}
\mathcal A_\alpha^{0\pm}\mathcal A_\beta^{\pm0}
&=
\frac12
\left(
\theta_\alpha\theta_\beta
+
\sin^2\theta\,\phi_\alpha\phi_\beta
\right)
\nonumber\\
&\quad
\pm\frac{i}{2}\sin\theta
\left(
\theta_\alpha\phi_\beta
-
\phi_\alpha\theta_\beta
\right).
\label{eq:app_product_0pm}
\end{align}
Taking the imaginary part gives
\begin{equation}
\operatorname{Im}
\left[
\mathcal A_\alpha^{0\pm}
\mathcal A_\beta^{\pm0}
\right]
=
\pm\frac{\sin\theta}{2}
\left(
\partial_{k_\alpha}\theta,\partial_{k_\beta}\phi
-
\partial_{k_\alpha}\phi,\partial_{k_\beta}\theta
\right).
\label{eq:app_im_product}
\end{equation}

It remains to express this result in terms of $\hat{\bm k}$. From
\begin{align}
k_x&=k\sin\theta\cos\phi \\
k_y&=k\sin\theta\sin\phi \\
k_z&=k\cos\theta,
\end{align}
one finds
\begin{align}
\partial_{k_x}\theta&=\frac{\cos\theta\cos\phi}{k} 
\quad, \quad \partial_{k_y}\theta=\frac{\cos\theta\sin\phi}{k} \\
\partial_{k_z}\theta&=-\frac{\sin\theta}{k} \\
\partial_{k_x}\phi&=-\frac{\sin\phi}{k\sin\theta} \quad, \quad
\partial_{k_y}\phi=\frac{\cos\phi}{k\sin\theta} \\
\partial_{k_z}\phi&=0.
\end{align}

Consequently,
\begin{equation}
\sin\theta
\left(
\partial_{k_\alpha}\theta\,\partial_{k_\beta}\phi
-
\partial_{k_\alpha}\phi\,\partial_{k_\beta}\theta
\right)
=
\frac{\epsilon_{\alpha\beta\mu}k_\mu}{k^3}.
\label{eq:app_spherical_identity}
\end{equation}

Substitution into Eq.~\eqref{eq:app_im_product} yields
\begin{equation}
\operatorname{Im}
\left[
\mathcal A_\alpha^{0\pm}
\mathcal A_\beta^{\pm0}
\right]
=
\pm
\frac{1}{2}
\frac{\epsilon_{\alpha\beta\mu}k_\mu}{k^3}.
\label{eq:app_berry_0pm_results}
\end{equation}

For completeness, consider the product involving the two dispersive helicity states. The spin-1 operators $S_x$ and $S_y$ connect only states with $\Delta m=\pm1$, while $S_z$ is diagonal. Hence,
\begin{equation}
\langle+1|S_i|-1\rangle=0,
\qquad i=x,y,z.
\end{equation}

It follows directly from Eq.~\eqref{eq:app_berry_connection} that
\begin{equation}
\mathcal A^{+-}=0,
\qquad
\mathcal A^{-+}=0.
\end{equation}

Therefore,
\begin{equation}
\operatorname{Im}
\left[
\mathcal A_\alpha^{+-}
\mathcal A_\beta^{-+}
\right]
=0.
\label{eq:app_berry_pm_results}
\end{equation}

Equations~\eqref{eq:app_berry_0pm_results} and
\eqref{eq:app_berry_pm_results} establish the Berry-connection products used in the main text.

\section{Hall response in the zero-temperature limit}
\label{app:zero_temperature}

At zero temperature, the Fermi--Dirac distribution reduces to a
Heaviside step function,
\begin{equation}
n_a(k)=\Theta(\mu-\varepsilon_a(k)),
\qquad
a=+1, 0, -1.
\end{equation}
Using the band energies
\begin{equation}
\varepsilon_\pm=\pm\hbar v_F k,
\qquad
\varepsilon_0=-ck^2,
\end{equation}
the Hall response can be written as
\begin{align}
\sigma_{xy}^{z,\mathrm{Hall}}
={}&
\frac{e^2}{24\pi^2g_L\mu_B}
\sum_{s=\pm 1}
\int_0^\infty\frac{dk}{k}
\left(3s v_F+\frac{c}{\hbar}k\right)
\nonumber\\[-1mm]
&\times
\left[
\Theta(\mu-\varepsilon_0)
-\Theta(\mu-\varepsilon_s)
\right].
\label{eq:app_sigma_Hall}
\end{align}
Separating the terms proportional to $v_F$ and $c$, we obtain
\begin{align}
\sigma_{xy}^{z,\mathrm{Hall}}
&=\sigma_L+\sigma_Q,
\end{align}
with
\begin{align}
\sigma_L
={}&
\frac{e^2v_F}{8\pi^2g_L\mu_B}
\int_0^\infty\frac{dk}{k}
\left[
\Theta(\mu-\varepsilon_-)
-\Theta(\mu-\varepsilon_+)
\right],
\label{eq:app_sigma_L}
\\[1mm]
\sigma_Q
={}&
\frac{e^2c}{24\pi^2\hbar g_L\mu_B}
\int_0^\infty dk
\Big[
2\Theta(\mu-\varepsilon_0)
\\[-1mm]
&\hspace{2.8cm}
-\Theta(\mu-\varepsilon_-)
-\Theta(\mu-\varepsilon_+)
\Big].
\label{eq:app_sigma_Q}
\end{align}

\subsection*{Linear contribution}

Since $\varepsilon_\pm=\pm\hbar v_Fk$, the combination entering
Eq.~\eqref{eq:app_sigma_L} is
\begin{align}
\Theta(\mu-\varepsilon_-)
-\Theta(\mu-\varepsilon_+)
&=
\Theta(\mu+\hbar v_Fk)
-\Theta(\mu-\hbar v_Fk).
\label{eq:app_theta_linear}
\end{align}
Introducing
\begin{equation}
k_F=\frac{|\mu|}{\hbar v_F},
\end{equation}
the two signs of $\mu$ lead to the same final step-function form. For
$\mu>0$,
\begin{align}
\Theta(\mu+\hbar v_Fk)&=1,
&
\Theta(\mu-\hbar v_Fk)
&=
\begin{cases}
1,&k<k_F,\\
0,&k>k_F,
\end{cases}
\end{align}
whereas for $\mu<0$,
\begin{align}
\Theta(\mu-\hbar v_Fk)&=0,
&
\Theta(\mu+\hbar v_Fk)
&=
\begin{cases}
0,&k<k_F,\\
1,&k>k_F.
\end{cases}
\end{align}
Thus, for $\mu\neq0$,
\begin{equation}
\Theta(\mu-\varepsilon_-)
-\Theta(\mu-\varepsilon_+)
=
\Theta(k-k_F).
\label{eq:app_theta_result}
\end{equation}

The logarithmic integral is ultraviolet sensitive, so we introduce the
cutoff $\Lambda$:
\begin{align}
\sigma_L
&=
\frac{e^2}{24\pi^2g_L\mu_B}3v_F
\int_{k_F}^{\Lambda}\frac{dk}{k}
\nonumber\\
&=
\frac{e^2v_F}{8\pi^2g_L\mu_B}
\ln\left(\frac{\Lambda}{k_F}\right)
\label{eq:app_sigma_L_final}
\end{align}


\subsection*{Quadratic contribution}

The quadratic contribution is obtained from
\begin{align}
\sigma_Q
={}&
\frac{e^2c}{24\pi^2\hbar g_L\mu_B}
\int_0^\infty dk
\Big[
2\Theta(\mu-\varepsilon_0)
\\[-1mm]
&\hspace{2.8cm}
-\Theta(\mu-\varepsilon_-)
-\Theta(\mu-\varepsilon_+)
\Big],
\label{eq:sigmaQ_integral}
\end{align}
where
\begin{equation}
\varepsilon_0=-ck^2,
\qquad
\varepsilon_\pm=\pm\hbar v_F k.
\end{equation}
Introducing the ultraviolet cutoff $\Lambda$ of the continuum model, we define
\begin{equation}
k_F=\frac{|\mu|}{\hbar v_F},
\qquad
k_0=\sqrt{\frac{|\mu|}{|c|}}.
\label{eq:kF_k0}
\end{equation}
We assume that the relevant momentum scales satisfy $k_F,k_0<\Lambda$.

We first consider the step functions associated with the linearly dispersing bands. For $\mu>0$,
\begin{equation}
\Theta(\mu-\varepsilon_-)
=
\Theta(\mu+\hbar v_F k)
=1,
\end{equation}
whereas
\begin{equation}
\Theta(\mu-\varepsilon_+)
=\Theta(\mu-\hbar v_F k)=
\begin{cases}
1, & k<k_F,\\
0, & k>k_F.
\end{cases}
\end{equation}
For $\mu<0$,
\begin{equation}
\Theta(\mu-\varepsilon_-)
=\Theta(\mu+\hbar v_F k)=
\begin{cases}
0, & k<k_F,\\
1, & k>k_F,
\end{cases}
\end{equation}
while
\begin{equation}
\Theta(\mu-\varepsilon_+)
=
\Theta(\mu-\hbar v_F k)
=0.
\end{equation}

The central band gives
\begin{equation}
\Theta(\mu-\varepsilon_0)
=
\Theta(\mu+ck^2),
\end{equation}
whose behavior depends on the relative signs of $\mu$ and $c$. For $\mu>0$,
\begin{equation}
\Theta(\mu+ck^2)
=
\begin{cases}
1, & c>0,\\[1mm]
\Theta(k_0-k), & c<0,
\end{cases}
\end{equation}
whereas for $\mu<0$,
\begin{equation}
\Theta(\mu+ck^2)
=
\begin{cases}
\Theta(k-k_0), & c>0,\\[1mm]
0, & c<0.
\end{cases}
\end{equation}

We can now combine the step functions. When $\mu c>0$, the resulting combination is
\begin{equation}
2\Theta(\mu-\varepsilon_0)
-\Theta(\mu-\varepsilon_-)
-\Theta(\mu-\varepsilon_+)
=
\operatorname{sgn}(c)\,\Theta(k-k_F).
\end{equation}
Consequently,
\begin{equation}
\begin{aligned}
\sigma_Q
&=
\frac{e^2c}{24\pi^2\hbar g_L\mu_B}
\operatorname{sgn}(c)
\int_{k_F}^{\Lambda}dk
&=
\frac{e^2|c|}{24\pi^2\hbar g_L\mu_B}
(\Lambda-k_F).
\end{aligned}
\label{eq:sigmaQ_same_sign}
\end{equation}

For $\mu c<0$, the two possible sign combinations have different step-function structures. If $\mu<0$ and $c>0$,
\begin{equation}
2\Theta(\mu-\varepsilon_0)
-\Theta(\mu-\varepsilon_-)
-\Theta(\mu-\varepsilon_+)
=
2\Theta(k-k_0)-\Theta(k-k_F),
\end{equation}
and therefore
\begin{equation}
\begin{aligned}
\int_0^\Lambda dk\,
\left[
2\Theta(k-k_0)-\Theta(k-k_F)
\right]
&=
2(\Lambda-k_0)-(\Lambda-k_F)
\\
&=
\Lambda+k_F-2k_0.
\end{aligned}
\end{equation}

If $\mu>0$ and $c<0$, one instead obtains
\begin{equation}
2\Theta(\mu-\varepsilon_0)
-\Theta(\mu-\varepsilon_-)
-\Theta(\mu-\varepsilon_+)
=
2\Theta(k_0-k)-1-\Theta(k_F-k),
\end{equation}
so that
\begin{equation}
\begin{aligned}
\int_0^\Lambda dk\,
\left[
2\Theta(k_0-k)-1-\Theta(k_F-k)
\right]
&=
2k_0-\Lambda-k_F
\\
&=
-(\Lambda+k_F-2k_0).
\end{aligned}
\end{equation}

Combining both sign combinations with the prefactor $c$, the result can be written compactly as
\begin{equation}
\sigma_Q
=
\frac{e^2|c|}{24\pi^2\hbar g_L\mu_B}
(\Lambda+k_F-2k_0),
\qquad
\mu c<0.
\label{eq:sigmaQ_opposite_sign}
\end{equation}

Thus, for $\mu\neq0$, the quadratic contribution is
\begin{equation}
\sigma_Q
=
\frac{e^2|c|}{24\pi^2\hbar g_L\mu_B}
\begin{cases}
\Lambda-k_F,
& \mu c>0,
\\[2mm]
\Lambda+k_F-2k_0,
& \mu c<0,
\end{cases}
\label{eq:sigmaQ_final}
\end{equation}
with
\begin{equation}
k_F=\frac{|\mu|}{\hbar v_F},
\qquad
k_0=\sqrt{\frac{|\mu|}{|c|}}.
\end{equation}
The different occupation patterns for the four combinations of signs of $\mu$ and $c$ therefore reduce to two cases according to the sign of $\mu c$. Equivalently, Eq.~\eqref{eq:sigmaQ_final} can be written in the compact form
\begin{equation}
\sigma_Q
=
\frac{e^2|c|}{24\pi^2\hbar g_L\mu_B}
\left[
\Lambda-k_F
+
2\Theta(-\mu c)
\left(k_F-k_0\right)
\right],
\end{equation}
where $\Theta$ denotes the Heaviside step function.

\section{Asymptotic scaling of the differential Hall response}
\label{app:scaling_D}

The finite-temperature differential Hall response receives contributions
from the linearly dispersing bands and the quadratic central band.
Although these sectors enter the response through the same microscopic
expression, their different dispersions give rise to distinct thermal
momentum scales. We first obtain an exact decomposition of the
finite-temperature response into the corresponding contributions and
then extract its asymptotic structure in the scaling limit
$t\to0$ at fixed $X=x/t$.

Starting from the dimensionless finite-temperature response,
\begin{equation}
\begin{aligned}
\bar{D}_{xy}^{z}(x,t)
={}&
\frac{1}{t}
\sum_{s=\pm1}
\int_0^\infty\frac{dq}{q}\,
s\left[3+s\,s_c q\right]\\
&\times
\left[
n_0(1-n_0)
-
n_s(1-n_s)
\right],
\end{aligned}
\label{eq:D_scaling_start}
\end{equation}
where
\begin{equation}
s_c=\operatorname{sgn}(c),
\quad
n_s=\frac{1}{e^{(sq-x)/t}+1},
\quad
n_0=\frac{1}{e^{(-s_cq^2-x)/t}+1}.
\end{equation}
It is convenient to introduce
\begin{equation}
h(z)
\equiv
\frac{e^z}{(1+e^z)^2}
=
\frac{1}{4\cosh^2(z/2)},
\label{eq:h_definition}
\end{equation}
so that
\begin{equation}
n_s(1-n_s)
=
h\left(\frac{s q-x}{t}\right),
\qquad
n_0(1-n_0)
=
h\left(\frac{-s_cq^2-x}{t}\right).
\end{equation}
We also define the scaling variable
\begin{equation}
X=\frac{x}{t}.
\label{eq:X_scaling}
\end{equation}

Using Eq.~\eqref{eq:h_definition}, the response can be separated
according to the two terms in the kernel,
\begin{equation}
\bar D_{xy}^{z}
=
\bar D_{(1)}
+
\bar D_{(2)},
\end{equation}
with
\begin{equation}
\bar D_{(1)}
=
\frac{3}{t}
\sum_{s=\pm1}s
\int_0^\infty\frac{dq}{q}
\left[
h\left(\frac{-s_cq^2-x}{t}\right)
-
h\left(\frac{s q-x}{t}\right)
\right],
\label{eq:D1}
\end{equation}
and
\begin{equation}
\bar D_{(2)}
=
\frac{s_c}{t}
\sum_{s=\pm1}
\int_0^\infty dq
\left[
h\left(\frac{-s_cq^2-x}{t}\right)
-
h\left(\frac{s q-x}{t}\right)
\right].
\label{eq:D2}
\end{equation}

The quadratic-band contribution in Eq.~\eqref{eq:D1} is independent
of $s$ and therefore vanishes upon summation.  Hence,
\begin{equation}
\bar D_{(1)}
=
-\frac{3}{t}
\sum_{s=\pm1}s
\int_0^\infty\frac{dq}{q}
h\left(\frac{s q-x}{t}\right).
\label{eq:D1_simplified}
\end{equation}
For the linearly dispersing bands, the thermal window is set by
$q\sim t$.  We therefore introduce
\begin{equation}
q=t y,
\qquad
\frac{dq}{q}=\frac{dy}{y},
\end{equation}
for which
\begin{equation}
\frac{s q-x}{t}=s y-X.
\end{equation}
Equation~\eqref{eq:D1_simplified} then becomes
\begin{equation}
\bar D_{(1)}
=
\frac{1}{t}\mathcal F(X),
\label{eq:D1_scaling}
\end{equation}
where
\begin{equation}
\mathcal F(X)
=
-3
\int_0^\infty\frac{dy}{y}
\left[
h(y-X)-h(-y-X)
\right].
\end{equation}
Thus, the contribution generated by the linear bands depends on $x$
and $t$ only through the ratio $X=x/t$ and scales as $t^{-1}$.

The second contribution in Eq.~\eqref{eq:D2} contains the quadratic
dispersion explicitly.  For this sector, the thermal window is
determined by
\begin{equation}
\frac{q^2}{t}\sim1,
\end{equation}
so that the appropriate scaling is
\begin{equation}
q=\sqrt{t}\,y,
\qquad
dq=\sqrt{t}\,dy.
\label{eq:q_sqrtt_scaling}
\end{equation}
The quadratic-band argument then becomes
\begin{equation}
\frac{-s_cq^2-x}{t}
=
-s_cy^2-X.
\end{equation}
Since this contribution is independent of $s$, the sum over the two
values of $s$ gives
\begin{equation}
2\int_0^\infty dq\,
h\left(\frac{-s_cq^2-x}{t}\right)
=
2\sqrt{t}
\int_0^\infty dy\,
h(-s_cy^2-X).
\end{equation}
After multiplication by the prefactor $s_c/t$, this gives
\begin{equation}
\bar D_{\rm Q}^{(0)}
=
\frac{1}{\sqrt{t}}\mathcal G(X),
\label{eq:DQ_scaling}
\end{equation}
where
\begin{equation}
\mathcal G(X)
=
2s_c
\int_0^\infty dy\,
h(-s_cy^2-X).
\end{equation}
Using the evenness of $h(z)$, this can equivalently be written as
\begin{equation}
\mathcal G(X)
=
\frac{s_c}{2}
\int_0^\infty dy\,
\operatorname{sech}^2
\left(
\frac{s_cy^2+X}{2}
\right).
\label{eq:G_scaling_sech}
\end{equation}

The remaining part of Eq.~\eqref{eq:D2} originates from the linearly
dispersing bands.  Using again $q=t y$, one obtains
\begin{equation}
-\sum_{s=\pm1}
\int_0^\infty dq\,
h\left(\frac{s q-x}{t}\right)
=
-t
\sum_{s=\pm1}
\int_0^\infty dy\,h(sy-X).
\end{equation}
This contribution is of order $t^0$ and defines the scaling function
\begin{equation}
\mathcal H(X)
=
-s_c
\int_0^\infty dy\,
\left[
h(y-X)+h(-y-X)
\right].
\end{equation}
Combining the two contributions, we obtain
\begin{equation}
\bar D_{xy}^{z}(x,t)
=
\frac{1}{t}\mathcal F(X)
+
\frac{1}{\sqrt{t}}\mathcal G(X)
+
\mathcal H(X),
\qquad
X=\frac{x}{t}.
\label{eq:D_complete_scaling}
\end{equation}
Equation~\eqref{eq:D_complete_scaling} establishes the scaling structure
directly from the original finite-temperature response.  In particular,
the first two terms are
\begin{equation}
\bar D_{xy}^{z}(x,t)
=
t^{-1}\mathcal F\left(\frac{x}{t}\right)
+
t^{-1/2}\mathcal G\left(\frac{x}{t}\right)
+\cdots .
\end{equation}

\bibliography{semiclassic,orbitronic,spin1_references,multifoldfermions}

\end{document}